\documentclass[12pt]{iopart}

\usepackage[utf8]{inputenc} 
\usepackage[T1]{fontenc} 
\usepackage[
  colorlinks=true,
  urlcolor=purple,
  citecolor=purple,
  filecolor=purple,
  linkcolor=purple
]{hyperref} 
\usepackage{amsopn, amssymb, graphicx, xcolor} 
\usepackage[subfolder]{gnuplottex} 
\usepackage{pifont}

\begin{document}

\newcommand\eqref[1]{\eref{#1}}
\makeatletter
\renewcommand\tableofcontents{\@starttoc{toc}}
\makeatother

\title{Analytic continuation over complex landscapes}

\author{Jaron Kent-Dobias and Jorge Kurchan}

\address{Laboratoire de Physique de l'Ecole Normale Supérieure, Paris, France}

\begin{abstract}
  In this paper we follow up the study of `complex complex landscapes'
  \cite{Kent-Dobias_2021_Complex}, rugged landscapes of many complex
  variables.  Unlike real landscapes, the classification of 
  saddles by index is trivial. Instead, the spectrum of fluctuations at stationary points determines their
  topological stability under analytic continuation of the
  theory. Topological changes, which occur at so-called Stokes points, proliferate among
 saddles with marginal (flat) directions and are suppressed
  otherwise. This gives a direct interpretation of the gap or `threshold' energy---which
  in the real case separates saddles from minima---as the level where the spectrum of
  the hessian matrix of stationary points develops a gap. This leads to different consequences
  for the analytic continuation of real landscapes with different structures:
  the global minima of `one step replica-symmetry broken' landscapes lie
  beyond a threshold, their hessians are gapped,  and are locally protected from Stokes points, whereas
  those of `many step replica-symmetry broken' have gapless hessians and
  Stokes points immediately proliferate.
  A new matrix ensemble is found, playing the role that GOE plays for real landscapes in determining
  the topological nature of saddles.
\end{abstract}

\maketitle

\tableofcontents

\section{Introduction}

Complex landscapes are functions of many variables having many minima
and, inevitably, many saddles of all indices (their number of unstable
directions). Optimization attempts to find the deepest minima, often
a difficult task.  For example, particles with a repulsive mutual potential
enclosed in a box will have many stable configurations, and we are asked to
find the one with lowest energy.

An aim of complexity studies is to classify these landscapes into
families having common properties. Two simplifications make the task
potentially tractable. The first is to consider the limit of many variables; in
the example of the particles, the limit of many particles, i.e. the
thermodynamic limit.  The
second simplification is of a more technical nature:  we often consider  functions that
contain some random element to them, and we study the ensemble average over that randomness.
The paradigm of this are spin-glasses, where the interactions are random, and
we are asked to find the ground state energy for typical samples.

Spin glass theory gave a surprise: random landscapes come in two kinds.
The first kind have a `threshold level' of energy, below which there are many
minima but almost no saddles, resulting in low minima that are separated by high barriers. The second have
all sorts of saddles all the way down to the lowest energy levels, and local
minima are separated by  barriers of sub-extensive energy height.  The latter are still
complex, but good solutions are easier to find.  This classification is closely
related to the structure of their replica trick solutions, the former being `one step replica-symmetry broken' and the latter being `many step replica-symmetry broken.'  Armed with this
solvable random example, it was easy to find non-random examples that behave
(at least approximately) in these two ways. For example,  sphere packings and the
travelling salesman problem belong to first and second classes, respectively.

What about the classification of systems whose variables are not real, but rather, complex?  Recalling
the Cauchy--Riemann conditions, one finds a difficulty: if our cost is,
say, the real part of a function of $N$ complex variables, in terms of the
corresponding $2N$ real variables it has only saddles of index $N$.  Even
worse: often not all saddles are equally interesting, so simply finding the
lowest is not usually what we need to do.  As it turns
out, there is a set of interesting questions to ask, as we describe below. For
each saddle, there is a `thimble' spanned by the lines along which the cost
function decreases. The way in which these thimbles fill the complex space is
crucial for many problems of analytic continuation, and is thus what we need to
study. The central role played by saddles in a real landscape, the `barriers',
is now played by the Stokes lines, by which thimbles exchange their properties.
Perhaps not surprisingly, the two classes of real landscapes described above
retain their significance in the complex case, but the distinction is now that while
in the first class the Stokes lines among the lowest minima are rare, in the
second class they proliferate.

In this paper we shall start from a many-variable integral of a real function,
and deform it in the many variable complex configuration space.  The landscape one faces
occupies the entirety of this space, and we shall see that this is an example where the
proliferation -- or lack of it -- of Stokes lines is the interesting quantity
in this context.

As for analytic continuation of physical theories: it is sometimes useful. Some theories
have a well-motivated Hamiltonian or action that nevertheless results in a
divergent partition function, and can only be properly defined by continuation
from a parameter regime where everything is well-defined
\cite{Witten_2011_Analytic}. Others result in oscillatory configuration space measures
that spoil the use of Monte Carlo or saddle point techniques, but can be
treated in a regime where the measure does not oscillate and the results
continued to the desired model \cite{Alexandru_2022_Complex}.

In any case, the nicest modern technique (which we will describe in some
detail) consists of deforming the configuration space integral into a complex configuration
space and then breaking it into pieces associated with stationary points of the
action. Each of these pieces, known as \emph{thimbles}, has wonderful
properties that guarantee convergence and prevent oscillations. Once such a
decomposition is made, analytic continuation is mostly easy, save for instances
where the thimbles interact, which must be accounted for.

When your action has a manageable set of stationary points, this process is
often tractable. However, many actions of interest are complex, having many
stationary points with no simple symmetry relating them, far too many to
individually track. Besides appearing in classical descriptions of structural
and spin glasses, complex landscapes have recently become important objects of
study in the computer science of machine learning, the condensed matter theory
of strange metals, and the high energy physics of black holes. What becomes of
analytic continuation under these conditions?

\section{Analytic continuation by thimble decomposition}
\label{sec:thimble.integration}

\subsection{Decomposition of the partition function into thimbles}
\label{subsec:thimble.decomposition}

Consider an action $\mathcal S$ defined on the (real) configuration space $\Omega$. A
typical calculation stems from a configuration space average of some observable
$\mathcal O$ of the form
\begin{equation} \label{eq:observable}
  \langle\mathcal O\rangle=\frac1Z\int_\Omega ds\,e^{-\beta\mathcal S(s)}\mathcal O(s)
\end{equation}
where the partition function $Z$ normalizes the average as
\begin{equation} \label{eq:partition.function}
  Z=\int_\Omega ds\,e^{-\beta\mathcal S(s)}
\end{equation}
Rather than focus on any specific observable, we will study the partition
function itself, since it exhibits the essential features.

\begin{figure}
  \hspace{5pc}
  \includegraphics{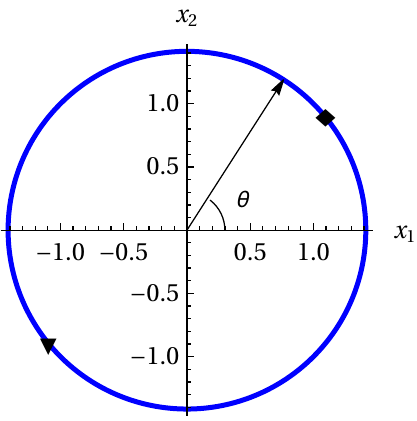}\hfill
  \includegraphics{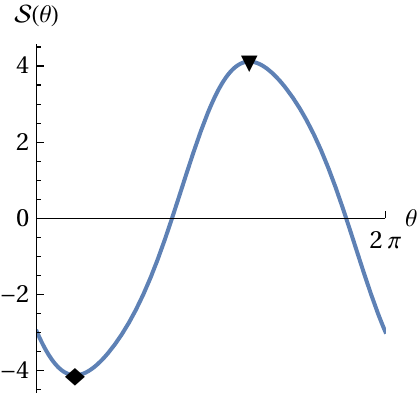}\hfill
  \includegraphics{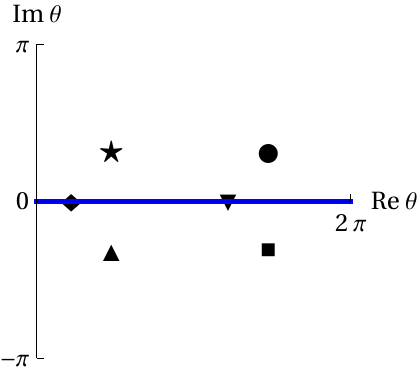}

  \caption{
    An example of a simple action and its stationary points. \textbf{Left:} The
    configuration space of the $N=2$ spherical (or circular) model, defined for
    $x\in\mathbb R^N$ restricted to the circle $N=x^Tx$. It can be parameterized
    by one angle $\theta=\arctan(x_2/x_1)$. Its natural complex extension takes
    instead $z\in\mathbb C^N$ restricted to the hyperbola
    $N=z^Tz=\|\operatorname{Re}z\|^2-\|\operatorname{Im}z\|^2$. The (now complex)
    angle $\theta$ is still a good parameterization of configuration space.
    \textbf{Center:} An action $\mathcal S$ for circular $3$-spin model,
    defined by $\mathcal
    S(z_1,z_2)=-1.051z_1^3-1.180z_1^2z_2-0.823z_1z_2^2-1.045z_2^3$, plotted as
    a function of $\theta$. \textbf{Right:} The stationary points of $\mathcal
    S$ in the complex-$\theta$ plane. In this example,
    $\Sigma=\{\mbox{\ding{117}},{\mbox{\ding{72}}},{\mbox{\ding{115}}},{\mbox{\ding{116}}},{\mbox{\ding{108}}},{\mbox{\ding{110}}}\}$
    and $\Sigma_0=\{\mbox{\ding{117}},{\mbox{\ding{116}}}\}$. Symmetries exist
    between the stationary points both as a result of the conjugation symmetry
    of $\mathcal S$, which produces the vertical reflection, and because in the
    pure 3-spin models $\mathcal S(-z)=-\mathcal S(z)$, which produces the
    simultaneous translation and inversion symmetry.
  } \label{fig:example.action}
\end{figure}

We've defined $Z$ in a way that suggests application in statistical mechanics,
but everything here is general: the action can be complex- or even
imaginary-valued, and $\Omega$ could be infinite-dimensional. In typical
contexts, $\Omega$ will be the euclidean real space $\mathbb R^N$ or some
subspace of this like the sphere $S^{N-1}$ (as in the $p$-spin spherical models
on which we will treat later). In this paper we will consider only analytic
continuation of the parameter $\beta$, but any other parameter would work
equally well, e.g., of some parameter inside the action. The action for real $\beta$ will have
some stationary points in the real configuration space, i.e., minima, maxima, saddles, and the set of those
points in $\Omega$ we will call $\Sigma_0$, the set of real stationary points.
An example action used throughout this section is shown in
Fig.~\ref{fig:example.action}.

In order to analytically continue \eref{eq:partition.function}, $\mathcal S$
must have an extension to a holomorphic function on a larger complex configuration
space $\tilde\Omega$ containing $\Omega$. In many cases this is accomplished by
noticing that the action is some sum or product of holomorphic
functions, e.g., polynomials, and replacing its real arguments with complex
ones. For $\mathbb R^N$ the complex configuration space $\tilde\Omega$ is $\mathbb
C^N$, while for the sphere $S^{N-1}$ it takes a little more effort. $S^{N-1}$
can be defined by all points $x\in\mathbb R^N$ such that $x^Tx=1$. A complex
extension of the sphere is made by extending this constraint: all points $z\in
\mathbb C^N$ such that $z^Tz=1$.  Both cases are complex manifolds, since they are defined by holomorphic constraints, and
therefore admit a hermitian metric and a symplectic structure. In the extended
complex configuration space, the action often has more stationary points. We'll
call $\Sigma$ the set of \emph{all} stationary points of the action, which
naturally contains the set of \emph{real} stationary points $\Sigma_0$.

\begin{figure}
  \hspace{5pc}
  \includegraphics{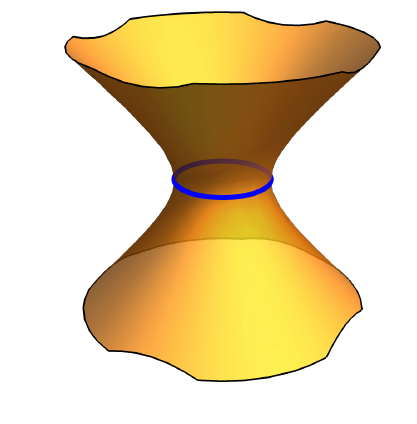}\hfill
  \includegraphics{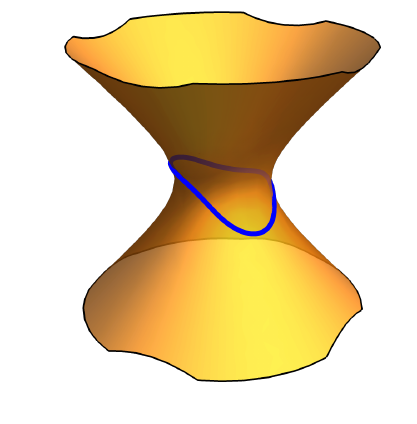}\hfill
  \includegraphics{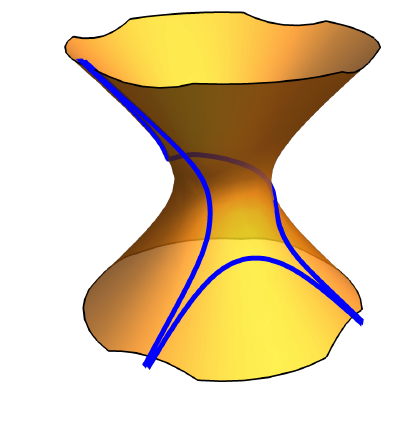}

  \hspace{5pc}
  \includegraphics{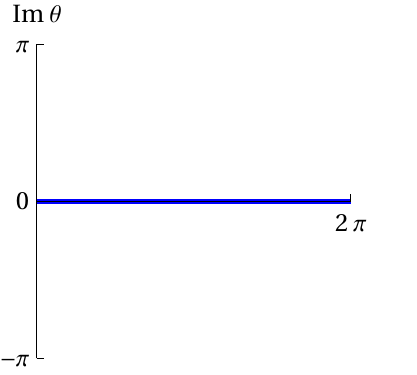}\hfill
  \includegraphics{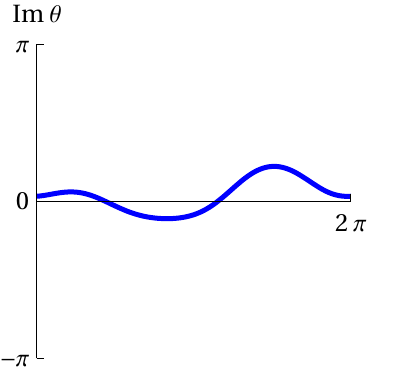}\hfill
  \includegraphics{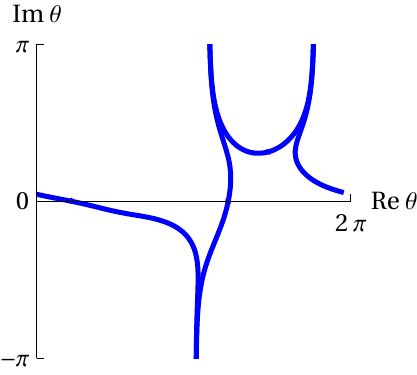}

  \caption{
    A schematic picture of the complex configuration space for the circular $p$-spin
    model and its standard integration contour. \textbf{Top:} For real variables,
    the model is a circle, and its analytic continuation is a kind of complex
    hyperbola, here shown schematically in three dimensions. \textbf{Bottom:}
    Since the real manifold (the circle) is one-dimensional, the complex
    manifold has one complex dimension, here parameterized by the angle
    $\theta$ on the circle. \textbf{Left:} The integration contour over the real configuration
    space of the circular model. \textbf{Center:} Complex analysis implies that the
    contour can be freely deformed without changing the value of the integral.
    \textbf{Right:} A funny deformation of the contour in which pieces have been
    pinched off to infinity. So long as no poles have been crossed, even this
    is legal.
  } \label{fig:contour.deformation}
\end{figure}

Assuming $\mathcal S$ is holomorphic (and that the configuration space $\Omega$ is
orientable, which is usually true) the integral in \eref{eq:partition.function}
can be considered an integral over a contour in the complex configuration space $\tilde\Omega$,
or
\begin{equation} \label{eq:contour.partition.function}
  Z=\oint_\Omega ds\,e^{-\beta\mathcal S(s)}
\end{equation}
For the moment this translation has only changed a symbol from
\eref{eq:partition.function}, but conceptually it is important: contour
integrals can have their contour freely deformed (under some constraints)
without changing their value. This means that we are free to choose a nicer
contour than our initial configuration space $\Omega$. This is illustrated in
Fig.~\ref{fig:contour.deformation}.

What properties are desirable for our contour? Consider the two motivations for performing analytic continuation cited
in the introduction: we
want our partition function to be well-defined, i.e., for the configuration space
integral to converge, and we want to avoid oscillations in the phase of the
integrand. The first condition, convergence, necessitates that the real part of
the action $\operatorname{Re}\beta\mathcal S$ be bounded from below, and that it
approach infinity in any limiting direction along the contour. The second,
constant phase, necessitates that the imaginary part of the action
$\operatorname{Im}\beta\mathcal S$ be constant.

Remarkably, there is an elegant recipe for accomplishing both these criteria at
once, courtesy of Picard--Lefschetz theory. For a more thorough review, see
\cite{Witten_2011_Analytic}. We will construct our deformed contour out
of a collection of pieces called \emph{thimbles}.
There is one thimble $\mathcal J_\sigma$ associated with each of the stationary
points $\sigma\in\Sigma$ of the action, and it is defined by all points that
approach the stationary point $s_\sigma$ under gradient descent on
$\operatorname{Re}\beta\mathcal S$: each thimble is the basin of attraction of a saddle.

Thimbles guarantee convergent integrals by construction: the value of
$\operatorname{Re}\beta\mathcal S$ is bounded from below on the thimble $\mathcal J_\sigma$ by its value
$\operatorname{Re}\beta\mathcal S(s_\sigma)$ at the stationary point,
since all other points on the thimble must descend to reach it. And, as we will
see in the following subsection, thimbles guarantee constant phase for the
integrand as well, a result of the underlying complex geometry of the problem.

\begin{figure}
  \hspace{5pc}
  \includegraphics{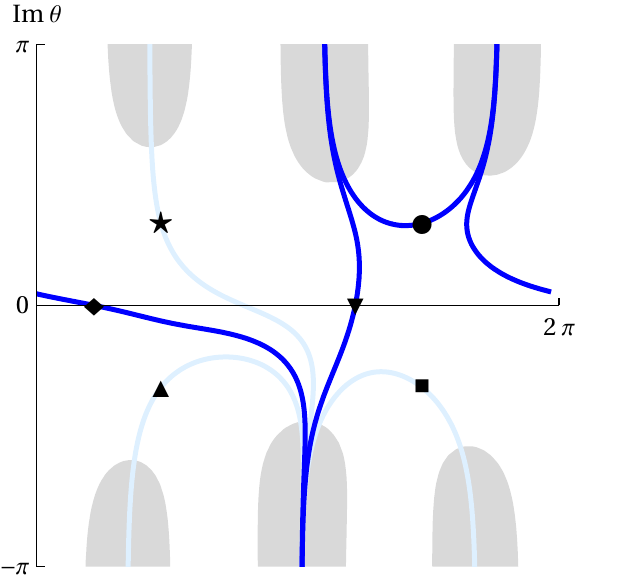}
  \hfill
  \includegraphics{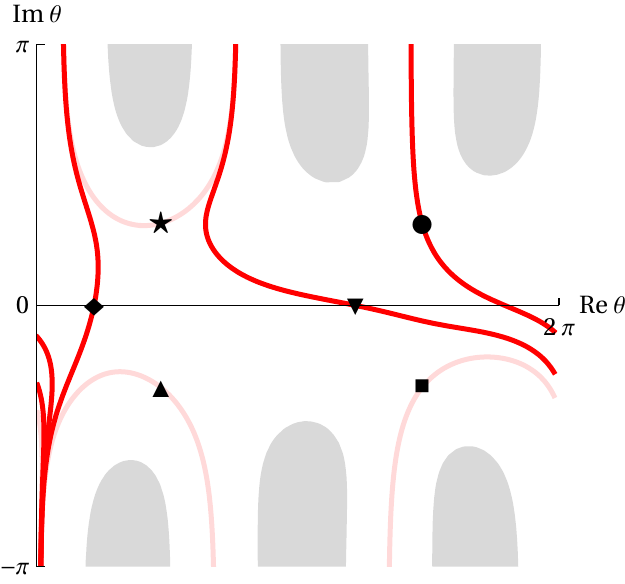}

  \caption{
    A demonstration of the rules of thimble homology. Both figures depict the
    complex-$\theta$ plane of action $\mathcal S$ featured in
    Fig.~\ref{fig:example.action} with $\arg\beta=0.4$. The black symbols lie
    on the stationary points of the action, and the grey regions depict the
    sets $\tilde\Omega_T$ of well-behaved regions at infinity (here $T=5$).
    \textbf{Left:} Lines show the thimbles of each stationary point. The
    thimbles necessary to recreate the cyclic path from left to right are
    darkly shaded, while those unnecessary for the task are lightly shaded.
    Notice that all thimbles come and go from the well-behaved regions.
    \textbf{Right:} Lines show the antithimbles of each stationary point.
    Notice that those of the stationary points involved in the contour (shaded
    darkly) all intersect the desired contour (the real axis), while those not
    involved do not intersect it.
  } \label{fig:thimble.homology}
\end{figure}

What thimbles are necessary to reproduce our original contour, $\Omega$? The
answer is, we need the minimal set which produces a contour between the same
places. Simply stated, if $\Omega=\mathbb R$ produced a configuration space integral
running along the real line from left to right, then our contour must likewise
go continuously from left to right, perhaps with detours to well-behaved
places at infinity (see Fig.~\ref{fig:thimble.homology}). The less simply stated versions follows.

Let $\tilde\Omega_T$ be the set of all points $z\in\tilde\Omega$ such that
$\operatorname{Re}\beta\mathcal S(z)\geq T$, where we will take $T$ to be a
very large number. $\tilde\Omega_T$ contains the parts of the manifold where it
is safe for any contour to end up if its integral is to converge, since
these are the places where the real part of the action is very large and the
real part of the integrand vanishes exponentially. The relative homology group
$H_N(\tilde\Omega,\tilde\Omega_T)$ describes the homology of $N$-dimensional cycles which begin
and end in $\Omega_T$, i.e., are well-behaved. Therefore, any well-behaved
cycle must represent an element of $H_N(\tilde\Omega,\tilde\Omega_T)$. In order
for our collection of thimbles to produce the correct contour, the composition
of the thimbles must represent the same element of this relative homology
group.

Each thimble represents an element of the relative homology, since each thimble
is a contour on which the real part of the action diverges at its extremes.
And, thankfully for us, Morse theory on our complex manifold $\tilde\Omega$
implies that the set of all thimbles produces a basis for this relative
homology group, and therefore any contour can be represented by some
composition of thimbles! There is even a systematic way to determine the
contribution from each thimble: for the stationary point $\sigma\in\Sigma$, let
$\mathcal K_\sigma$ be its \emph{antithimble}, defined by all points brought to
$s_\sigma$ by gradient \emph{ascent} (and representing an element of the
relative homology group $H_N(\tilde\Omega,\tilde\Omega_{-T})$). Then each
thimble $\mathcal J_\sigma$ contributes to the contour with a weight given by
its intersection pairing $n_\sigma=\langle\mathcal C,\mathcal K_\sigma\rangle$.

\begin{figure}
  \hspace{5pc}
  \includegraphics{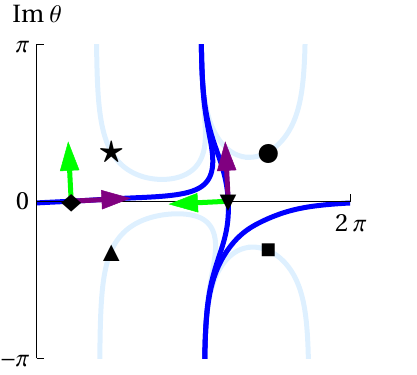}\hfill
  \includegraphics{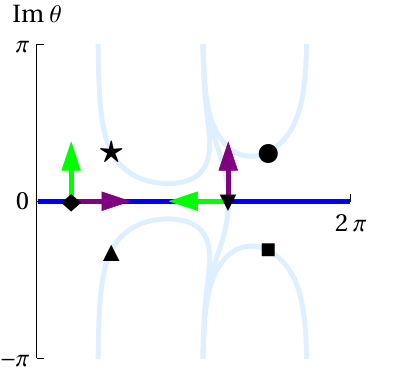}\hfill
  \includegraphics{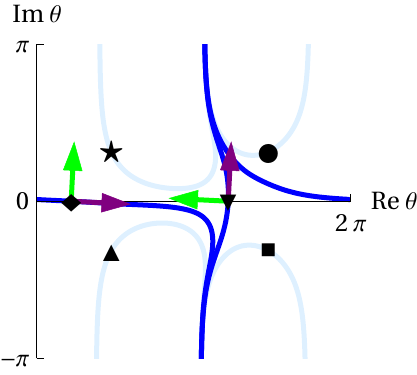}

  \caption{
    The behavior of thimble contours near $\arg\beta=0$ for real actions. In all
    pictures, green arrows depict a canonical orientation of the thimbles
    relative to the real axis, while purple arrows show the direction of
    integration implied by the orientation. \textbf{Left:} $\arg\beta=-0.1$. To
    progress from left to right, one must follow the thimble from the minimum
    $\mbox{\ding{117}}$ in the direction implied by its orientation, and then
    follow the thimble from the maximum ${\mbox{\ding{116}}}$ \emph{against} the
    direction implied by its orientation, from top to bottom. Therefore,
    $\mathcal C=\mathcal J_{\mbox{\ding{117}}}-\mathcal J_{\mbox{\ding{116}}}$.
    \textbf{Center:} $\arg\beta=0$. Here the thimble of the minimum covers
    almost all of the real axis, reducing the problem to the real configuration space
    integral. This is also a Stokes point. \textbf{Right:} $\arg\beta=0.1$. Here, one follows the thimble of
    the minimum from left to right again, but now follows that of the maximum
    in the direction implied by its orientation, from bottom to top. Therefore,
    $\mathcal C=\mathcal J_{\mbox{\ding{117}}}+\mathcal J_{\mbox{\ding{116}}}$.
  } \label{fig:thimble.orientation}
\end{figure}

\begin{figure}
  \hspace{5pc}
  \includegraphics{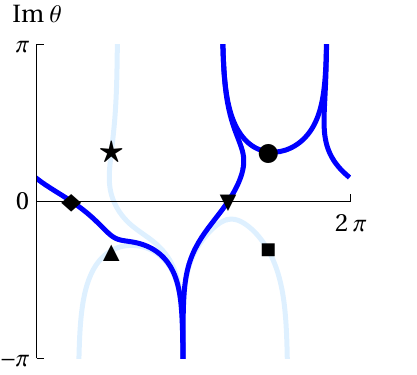}\hfill
  \includegraphics{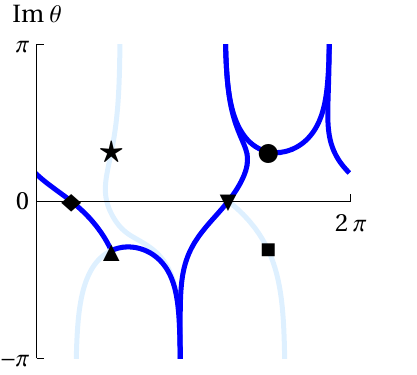}\hfill
  \includegraphics{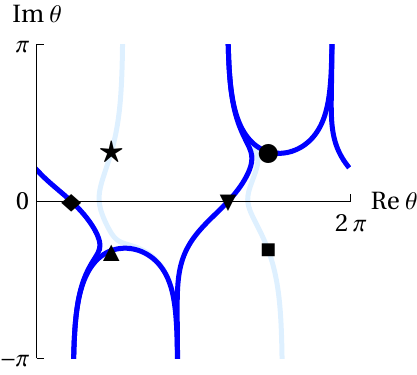}

  \caption{
    An example of a Stokes point in the continuation of the configuration space
    integral involving the action $\mathcal S$ featured in
    Fig.~\ref{fig:example.action}. \textbf{Left:} $\arg\beta=1.176$. The collection of
    thimbles necessary to progress around from left to right, highlighted in a
    darker color, is the same as it was in Fig.~\ref{fig:thimble.homology}.
    \textbf{Center:} $\arg\beta=1.336$. The thimble $\mathcal J_{\mbox{\ding{117}}}$
    intersects the stationary point ${\mbox{\ding{115}}}$ and its thimble, leading
    to a situation where the contour is not easily defined using thimbles. This
    is a Stokes point. \textbf{Right:} $\arg\beta=1.496$. The Stokes point has passed,
    and the collection of thimbles necessary to produce the path has changed:
    now $\mathcal J_{\mbox{\ding{115}}}$ must be included. Notice that in this
    figure, because of the symmetry of the pure models, the thimble $\mathcal
    J_{\mbox{\ding{110}}}$ also experiences a Stokes point, but this does not result
    in a change to the contour involving that thimble.
  } \label{fig:1d.stokes}
\end{figure}

With these tools in hands, we can finally write the partition function as a sum
over contributions from each thimble, or
\begin{equation} \label{eq:thimble.integral}
  Z=\sum_{\sigma\in\Sigma}n_\sigma\oint_{\mathcal J_\sigma}ds\,e^{-\beta\mathcal S(s)}.
\end{equation}
Under analytic continuation, the form of \eref{eq:thimble.integral}
generically persists. When the relative homology of the thimbles is unchanged
by the continuation, the integer weights are likewise unchanged, and one can
therefore use the knowledge of these weights in one regime to compute the
partition function in the other. However, their relative homology can change,
and when this happens the integer weights can be traded between stationary
points. These trades occur when two thimbles intersect, or alternatively when
one stationary point lies in the gradient descent of another. These places are
called \emph{Stokes points}, and the gradient descent trajectories that join
two stationary points are called \emph{Stokes lines}. An example of this
behavior can be seen in Fig.~\ref{fig:1d.stokes}.

The prevalence (or not) of Stokes points in a given continuation, and whether
those that do appear affect the weights of stationary points of interest, is a
concern for the analytic continuation of theories. If they do not occur or
occur order-one times, one could reasonably hope to perform such a procedure.
If they occur exponentially often in the system size, there is little hope of
keeping track of the resulting weights, and analytic continuation is intractable.

\subsection{Gradient flow}
\label{subsec:thimble.flow}

The `dynamics' describing thimbles is defined by gradient descent on the real
part of the action, with a given thimble incorporating all trajectories which
asymptotically flow to its associated stationary point. Since our configuration space
is not necessary flat (as for the \emph{spherical} $p$-spin models), we will
have to do a bit of differential geometry to work out the form of the flow. Gradient
descent on a complex manifold is given by
\begin{equation} \label{eq:flow.coordinate.free}
  \dot s
  =-\operatorname{grad}\operatorname{Re}\beta\mathcal S
  =-\left(\frac\partial{\partial s^*}\operatorname{Re}\beta\mathcal S\right)^\sharp
  =-\frac{\beta^*}2\frac{\partial\mathcal S^*}{\partial s^*}g^{-1}\frac\partial{\partial s}
\end{equation}
where $g$ is the metric and
$\partial\mathcal S/\partial s^*=0$ because the action is holomorphic. If the complex configuration space is $\mathbb C^N$ and the
metric is diagonal, this means that the flow is proportional to the conjugate
of the gradient, or $\dot s\propto-\beta^*(\partial\mathcal S/\partial s)^*$.

In the case we will consider here (namely, that of the spherical models), it
will be more convenient to work in terms of coordinates in a flat embedding
space than in terms of local coordinates in the curved space, e.g., in terms of
$z\in\mathbb C^N$ instead of $s\in S^{N-1}$. Let $z:\tilde\Omega\to\mathbb C^N$
be an embedding of complex configuration space into complex euclidean space. The
dynamics in the embedding space is given by
\begin{equation}\label{eq:flow.raw}
  \dot z
  =-\frac{\beta^*}2\frac{\partial\mathcal S^*}{\partial z^*}(Dz)^* g^{-1}(Dz)^T\frac\partial{\partial z}
\end{equation}
where $Dz=\partial z/\partial s$ is the Jacobian of the embedding.
The embedding induces a metric on $\tilde\Omega$ by $g=(Dz)^\dagger Dz$.
Writing $\partial=\partial/\partial z$, this gives
\begin{equation} \label{eq:flow}
  \dot z=-\frac{\beta^*}2(\partial\mathcal S)^\dagger(Dz)^*[(Dz)^\dagger(Dz)]^{-1}(Dz)^T
  =-\frac12(\partial \mathcal S)^\dagger P
\end{equation}
which is nothing but the projection of $(\partial\mathcal S)^*$ into the
tangent space of the manifold, with the projection operator
$P=(Dz)^*[(Dz)^\dagger(Dz)]^{-1}(Dz)^T$. Note that $P$ is hermitian.

Though the projection operator can be derived for any particular manifold by
defining a coordinate system and computing it with the above definition, for
simple manifolds like the sphere it can be guessed easily enough, as the unique
hermitian operator that projects out the direction normal to the surface. For
the sphere, this is
\begin{equation}
  P=I-\frac{zz^\dagger}{|z|^2}
\end{equation}
One can
quickly verify that this operator indeed projects the dynamics onto the
manifold: the vector perpendicular to the manifold at any point $z$ is given by
$\partial(z^Tz)=z$, and $Pz=z-z|z|^2/|z|^2=0$. For any vector $u$ perpendicular
to $z$, i.e., $z^\dagger u=0$, $Pu=u$, the identity.

\begin{figure}
  \hspace{5pc}
  \hfill
  \includegraphics{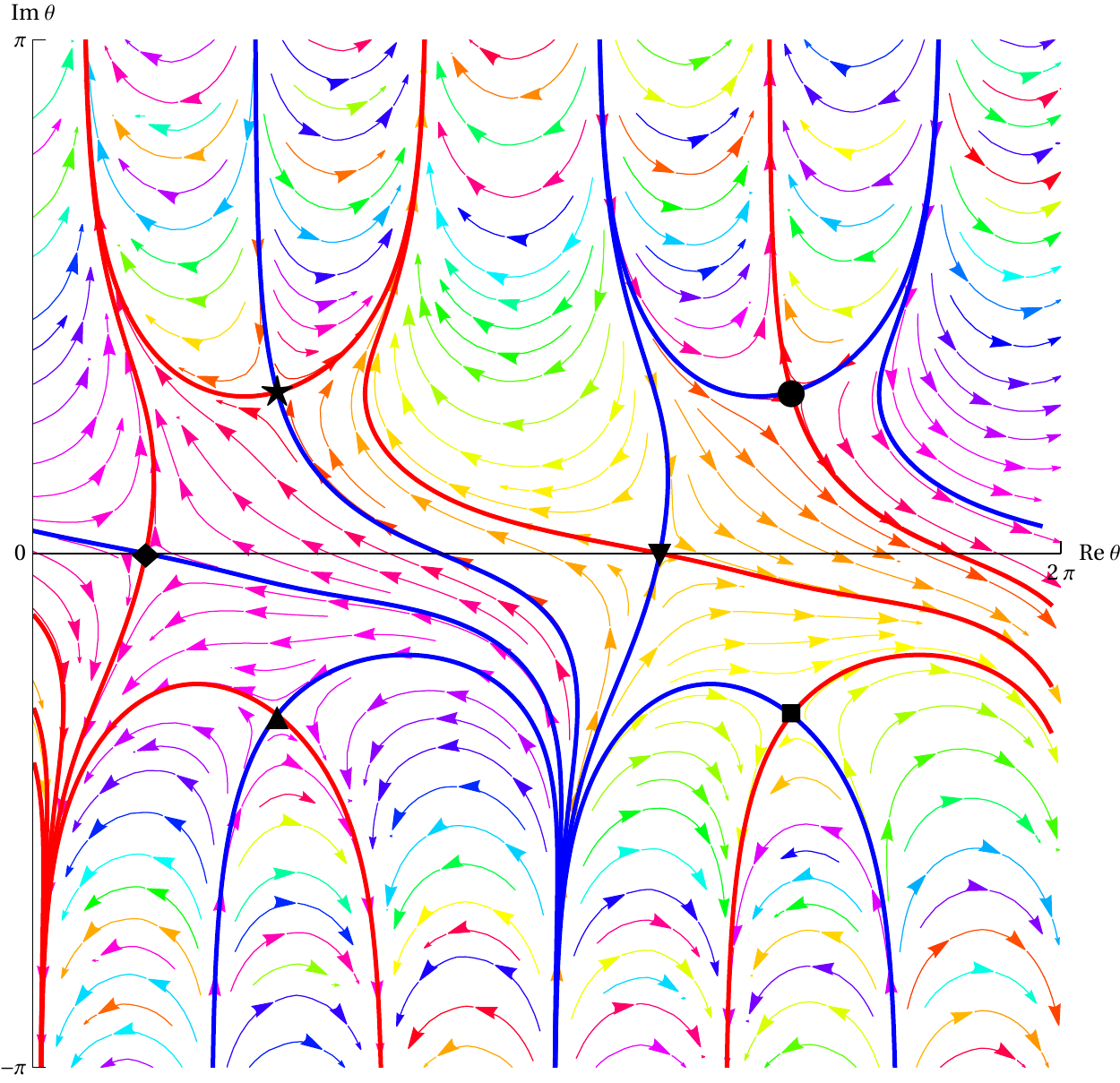}

  \caption{Example of gradient descent flow on the action $\mathcal S$ featured
    in Fig.~\ref{fig:example.action} in the complex-$\theta$ plane, with
    $\arg\beta=0.4$. Symbols denote the stationary points, while thick blue and
    red lines depict the thimbles and antithimbles, respectively. Streamlines
    of the flow equations are plotted in a color set by their value of
    $\operatorname{Im}\beta\mathcal S$; notice that the color is constant along
    each streamline.
  } \label{fig:flow.example}
\end{figure}

Gradient descent on $\operatorname{Re}\beta\mathcal S$ is equivalent to
Hamiltonian dynamics with the Hamiltonian $\operatorname{Im}\beta\mathcal S$
and conjugate coordinates given by the real and imaginary parts of each complex
coordinate. This is because $(\tilde\Omega, g)$ is a Kähler manifold and therefore admits
a symplectic structure, but it can be shown that the flow conserves the imaginary action
using \eref{eq:flow} and the
holomorphic property of $\mathcal S$:
\begin{equation}
  \eqalign{
    \frac d{dt}\operatorname{Im}\beta\mathcal S
    &=\dot z\partial\operatorname{Im}\beta\mathcal S+\dot z^*\partial^*\operatorname{Im}\beta\mathcal S \\
    &=\frac i4\left(
      (\beta\partial \mathcal S)^\dagger P\beta\partial\mathcal S-(\beta\partial\mathcal S)^TP^*(\beta\partial\mathcal S)^*
    \right) \\
    &=\frac{i|\beta|^2}4\left(
      (\partial\mathcal S)^\dagger P\partial\mathcal S-[(\partial\mathcal S)^\dagger P\partial\mathcal S]^*
    \right) \\
    &=\frac{i|\beta|^2}4\left(
      \|\partial\mathcal S\|^2-(\|\partial\mathcal S\|^*)^2
    \right)=0.
  }
\end{equation}
where $\|v\|^2=v^\dagger Pv$ is the norm of a complex vector $v$ in the tangent space of the manifold.
The flow of the action takes a
simple form:
\begin{equation}
  \dot{\mathcal S}
  =\dot z\partial\mathcal S
  =-\frac{\beta^*}2(\partial\mathcal S)^\dagger P\partial\mathcal S
  =-\frac{\beta^*}2\|\partial\mathcal S\|^2.
\end{equation}
In the complex-$\mathcal S$ plane, dynamics is occurs along straight lines in
a direction set by the argument of $\beta$.

\subsection{The conditions for Stokes points}
\label{subsec:stokes.conditions}

As we have seen, gradient descent on the real part of the action results in a
flow that preserves the imaginary part of the action.
Stokes lines, when they manifest, are topologically persistent so long as this
conservation is respected: if a Stokes line connects two stationary points and
the action is smoothly modified under the constraint that the imaginary parts
of the two stationary points is held equal, the Stokes line will continue to
connect them so long as the flow of a third stationary point does not sever
their connection, i.e., so long as there is not a topological change in the flow. This implies that, despite being relatively low-dimensional
surfaces of codimension $N$, thimble connections are found with a
codimension one tuning of parameters, modulo the topological adjacency
requirement. This means that, though not present in generic cases, Stokes points generically appear when a
dimension-one curve is followed in parameter space.

Not all Stokes points result in the exchange of weight between thimbles.
Examining Fig.~\ref{fig:1d.stokes} again, notice that the thimbles $\mathcal
J_{\mbox{\ding{110}}}$ and $\mathcal J_{\mbox{\ding{116}}}$ also experience a
Stokes point, but this does not result in a change to the contour involving those
thimbles. This is because the integer weights can only be modified when a thimble
that has some nonzero weight is downstream on the gradient descent flow, and
therefore a necessary condition for a meaningful change in the thimble
decomposition involving two stationary points $\sigma$ and $\tau$ where
$n_\sigma\neq0$ and $n_\tau=0$ is for $\operatorname{Re}\beta\mathcal
S(s_\sigma)<\operatorname{Re}\beta\mathcal S(s_\tau)$.

\begin{figure}
  \hspace{5pc}
  \hfill
  \includegraphics{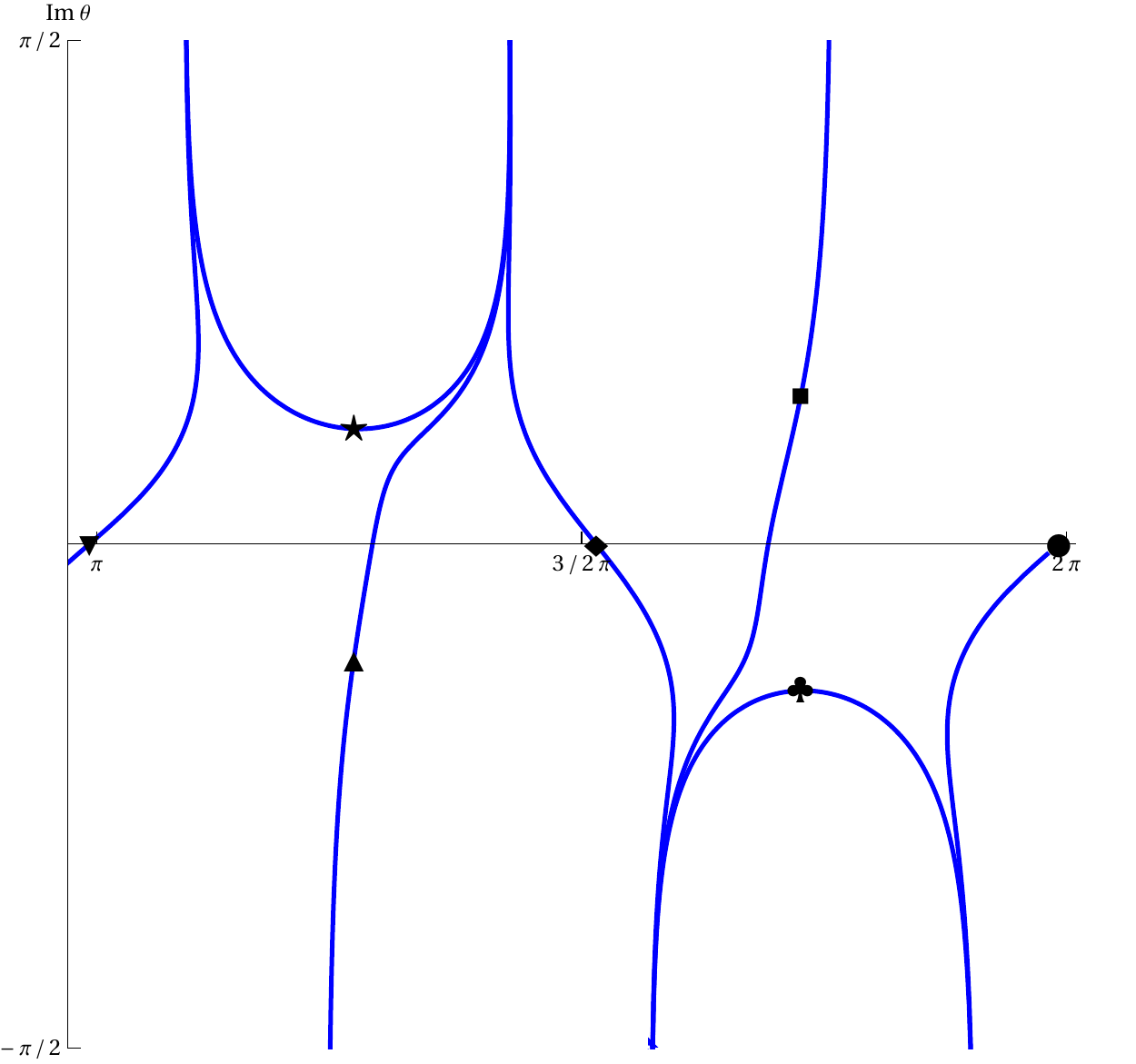}
  \caption{
  Some thimbles of the circular 6-spin model, where the argument of $\beta$ has
  been chosen such that the imaginary parts of the action at the stationary
  points $\clubsuit$ and ${\mbox{\ding{115}}}$ are exactly the same (and, as a
  result of conjugation symmetry, the points ${\mbox{\ding{72}}}$ and ${\mbox{\ding{110}}}$).
  } \label{fig:4.spin}
\end{figure}

\begin{figure}
  \hspace{5pc}
  \hfill
  \includegraphics{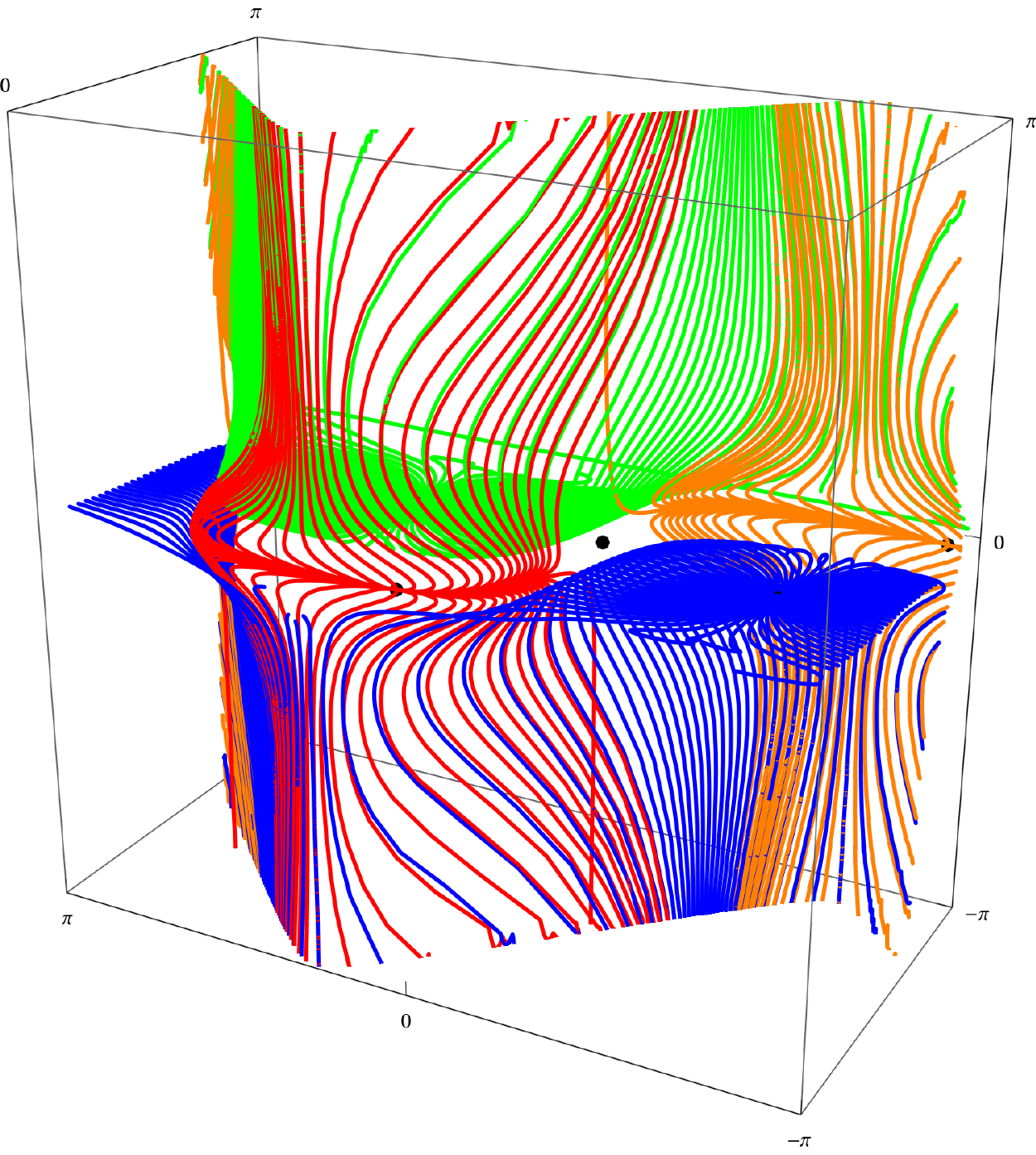}
  \caption{
    Thimbles of the $N=3$ spherical 2-spin model projected into the
    $\operatorname{Re}\theta$, $\operatorname{Re}\phi$,
    $\operatorname{Im}\theta$ space. The blue and green lines trace gradient
    descent of the two minima, while the red and orange lines trace those of
    the two saddles. The location of the maxima are marked as points, but their
    thimbles are not shown.
  } \label{fig:3d.thimbles}
\end{figure}

Another necessary condition for the existence of a Stokes line between two stationary
points is for those points to have the same imaginary action. However, this is
not a sufficient condition. This can be seen in Fig.~\ref{fig:4.spin}, which
shows the thimbles of the circular 6-spin model. The argument of $\beta$ has
been chosen such that the stationary points marked by $\clubsuit$ and
${\mbox{\ding{115}}}$ have exactly the same imaginary energy, and yet they do
not share a thimble.
This is because these stationary points are not adjacent: they are separated
from each other by the thimbles of other stationary points. This is a
consistent story in one complex dimension, since the codimension of the
thimbles is one, and thimbles can divide space into regions. However, in
higher dimensions thimbles do not have a codimension high enough to divide space
into regions.
Nonetheless, thimble intersections
are still governed by a requirement for adjacency. Fig.~\ref{fig:3d.thimbles}
shows a projection of the thimbles of an $N=3$ 2-spin model, which is defined
on the sphere. Because of an inversion symmetry of the model, stationary points
on opposite sides of the sphere have identical energies, and therefore also
share the same imaginary energy. However, their thimbles (blue and green in the
figure) do not intersect. Here, they could not possibly intersect, since the
real parts of their energy are also the same, and upward flow could therefore
not connect them.

Determining whether stationary points are adjacent in this sense is a difficult problem,
known as the global connection problem \cite{Howls_1997_Hyperasymptotics}. It
is also difficult for us to reason rigorously about the properties of
stationary point adjacency. However, we have a coarse argument for why, in
generic cases with random actions, one should expect the typical number of adjacent
stationary points to scale algebraically with dimension. First, notice that in
order for two stationary points to be eligible to share a Stokes point, their thimbles
must approach the same `good' region of complex configuration space. This is because weight is traded at Stokes points when a facet of
one thimble flops over another between good regions. Therefore, one can draw
conclusions about the number of stationary points eligible for a Stokes point
with a given stationary point by examining the connectivity of the `good'
regions.

In the one-dimensional examples above, the `good regions' for contours are
zero-dimensional, making their topology discrete. However, in a $D$-dimensional
case, these regions are $D-1$ dimensional, and their topology is richer.
Slices of thimbles evaluated at constant `height' as measured by the real part of the
action are topologically $D-1$ spheres. These slices are known as the \emph{vanishing cycles} of the thimble. At the extremal reaches of the configuration
space manifold, these spherical slices form a mesh, sharing sections of their
boundary with the slices of other thimbles and covering the extremal reaches
like a net. Without some special symmetry to produce vertices in this mesh
where many thimbles meet, such a mesh generally involves order $D$ boundaries
coming together in a given place. Considering the number of faces on a given
extremal slice should also be roughly linear in $D$, one expects something like
quadratic growth with $D$ of eligible neighbors, something which gives a rough sense of locality in Stokes point interactions.

\subsection{The structure of stationary points}
\label{subsec:stationary.hessian}

The shape of each thimble in the vicinity of its stationary point can be
described using an analysis of the hessian of the real part of the action at
the stationary point. Here we'll review some general properties of this
hessian, which has rich structure because the action is holomorphic.

Writing down the hessian using the complex geometry of the previous section
would be arduous. Luckily, we are only interested in the hessian at
stationary points, and our manifolds of interest are constraint surfaces.
These two facts allow us to find the hessian at stationary points using a
simpler technique, that of Lagrange multipliers.

Suppose that our complex manifold $\tilde\Omega$ is defined by all points
$z\in\mathbb C^N$ such that $g(z)=0$ for some holomorphic function $g$. In the
case of the spherical models, $g(z)=\frac12(z^Tz-N)$. Introducing the Lagrange
multiplier $\mu$, we define the constrained action
\begin{equation}
  \tilde\mathcal S(z)=\mathcal S(z)-\mu g(z)
\end{equation}
The condition for a stationary point is that $\partial\tilde\mathcal S=0$. This implies that, at any stationary point,
$\partial\mathcal S=\mu\partial g$. In particular, if $\partial g^T\partial g\neq0$, we find the value for the Lagrange multiplier $\mu$ as
\begin{equation} \label{eq:multiplier}
  \mu=\frac{\partial g^T\partial\mathcal S}{\partial g^T\partial g}
\end{equation}
As a condition for a stationary point, this can be intuited as projecting out
the normal to the constraint surface $\partial g$ from the gradient of the
unconstrained action. It implies that the hessian with respect to the complex
embedding coordinate $z$ at any stationary point is
\begin{equation} \label{eq:complex.hessian}
  \operatorname{Hess}\mathcal S
  =\partial\partial\tilde\mathcal S
  =\partial\partial\mathcal S-\frac{\partial g^T\partial\mathcal S}{\partial g^T\partial g}\partial\partial g
\end{equation}
In practice one must neglect the directions normal to the constraint surface by
projecting them out using $P$ from the previous section, i.e.,
$P\operatorname{Hess}\mathcal SP^T$. For notational simplicity we will not
include this here.

In order to describe the structure of thimbles, one must study the hessian of
$\operatorname{Re}\beta\mathcal S$, since it is the upward directions in the
flow on the real action in the vicinity of stationary points which define them.  We first
pose the problem as one
of $2N$ real variables $x,y\in\mathbb R^N$ with $z=x+iy$. The hessian of the
real part of the action with respect to these real variables is
\begin{equation} \label{eq:real.hessian}
  \operatorname{Hess}_{x,y}\operatorname{Re}\beta\mathcal S
  =\left[\matrix{
    \partial_x\partial_x\operatorname{Re}\beta\tilde\mathcal S &
    \partial_y\partial_x\operatorname{Re}\beta\tilde\mathcal S \cr
    \partial_x\partial_y\operatorname{Re}\beta\tilde\mathcal S &
    \partial_y\partial_y\operatorname{Re}\beta\tilde\mathcal S
  }\right]
\end{equation}
This can be simplified using the fact that the action is holomorphic, which
means that it obeys the Cauchy--Riemann equations
\begin{equation}
  \partial_x\operatorname{Re}\tilde\mathcal S=\partial_y\operatorname{Im}\tilde\mathcal S
  \qquad
  \partial_y\operatorname{Re}\tilde\mathcal S=-\partial_x\operatorname{Im}\tilde\mathcal S
\end{equation}
Using these relationships alongside the Wirtinger derivative
$\partial\equiv\frac12(\partial_x-i\partial_y)$ allows the order of the
derivatives and the real or imaginary parts to be commuted, with
\begin{equation}
  \eqalign{
    \partial_x\operatorname{Re}\tilde\mathcal S=\operatorname{Re}\partial\tilde\mathcal S
    \qquad
    \partial_y\operatorname{Re}\tilde\mathcal S=-\operatorname{Im}\partial\tilde\mathcal S \\
    \partial_x\operatorname{Im}\tilde\mathcal S=\operatorname{Im}\partial\tilde\mathcal S
    \qquad
    \partial_y\operatorname{Im}\tilde\mathcal S=\operatorname{Re}\partial\tilde\mathcal S
  }
\end{equation}
Using these relationships, the hessian \eref{eq:real.hessian} can be written in
the more manifestly complex way
\begin{equation}
  \eqalign{
    \operatorname{Hess}_{x,y}\operatorname{Re}\beta\mathcal S
    &=\left[\matrix{
      \hphantom{-}\operatorname{Re}\beta\partial\partial\tilde\mathcal S &
      -\operatorname{Im}\beta\partial\partial\tilde\mathcal S \cr
      -\operatorname{Im}\beta\partial\partial\tilde\mathcal S &
      -\operatorname{Re}\beta\partial\partial\tilde\mathcal S
    }\right] \\
    &=\left[\matrix{
        \hphantom{-}\operatorname{Re}\beta\operatorname{Hess}\mathcal S &
      -\operatorname{Im}\beta\operatorname{Hess}\mathcal S \cr
      -\operatorname{Im}\beta\operatorname{Hess}\mathcal S &
      -\operatorname{Re}\beta\operatorname{Hess}\mathcal S
    }\right]
  }
\end{equation}
where $\operatorname{Hess}\mathcal S$ is the hessian with respect to $z$ given
in \eqref{eq:complex.hessian}.

The eigenvalues and eigenvectors of the hessian are important for evaluating
thimble integrals, because those associated with upward directions provide a
local basis for the surface of the thimble. Suppose that $v_x,v_y\in\mathbb
R^N$ are such that
\begin{equation}
  (\operatorname{Hess}_{x,y}\operatorname{Re}\beta\mathcal S)\left[\matrix{v_x \cr v_y}\right]
  =\lambda\left[\matrix{v_x \cr v_y}\right]
\end{equation}
where the eigenvalue $\lambda$ must be real because the hessian is real symmetric. The problem can be put into a more obviously complex form by a change of basis. Writing $v=v_x+iv_y$, we find
\begin{equation}
  \eqalign{
    &\left[\matrix{0&(i\beta\operatorname{Hess}\mathcal S)^*\cr i\beta\operatorname{Hess}\mathcal S&0}\right]
    \left[\matrix{v \cr iv^*}\right]\\
    &\qquad=\left[\matrix{1&i\cr i&1}\right]
    (\operatorname{Hess}_{x,y}\operatorname{Re}\beta\mathcal S)
    \left[\matrix{1&i\cr i&1}\right]^{-1}
    \left[\matrix{1&i\cr i&1}\right]
    \left[\matrix{v_x \cr v_y}\right] \\
    &\qquad=\lambda\left[\matrix{1&i\cr i&1}\right]\left[\matrix{v_x \cr v_y}\right]
    =\lambda\left[\matrix{v \cr iv^*}\right]
  }
\end{equation}
It therefore follows that the eigenvalues and vectors of the real hessian satisfy the equation
\begin{equation} \label{eq:generalized.eigenproblem}
  \beta\operatorname{Hess}\mathcal S v=\lambda v^*
\end{equation}
a sort of generalized
eigenvalue problem whose solutions are called the \emph{Takagi vectors} of $\operatorname{Hess}\mathcal S$ \cite{Takagi_1924_On}. If we did not know the eigenvalues were real, we could
still see it from the second implied equation,
$(\beta\operatorname{Hess}\mathcal S)^*v^*=\lambda v$, which is the conjugate
of the first if $\lambda^*=\lambda$.

Something hidden in the structure of the real hessian but more clear
in its complex form is that each eigenvalue comes in a pair, since
\begin{equation}
  \beta\operatorname{Hess}\mathcal S(iv)=i\lambda v^*=-\lambda(iv)
\end{equation}
Therefore, if $\lambda$ satisfies \eqref{eq:generalized.eigenproblem} with Takagi vector $v$,
than so does $-\lambda$ with associated Takagi vector $iv$, rotated in the complex
plane. It follows that each stationary point has an equal number of descending
and ascending directions, i.e., the index of each stationary point is $N$. For
a stationary point in a real problem this might seem strange, because there are
clear differences between minima, maxima, and saddles of different index.
However, for  such a stationary point, its $N$ real Takagi vectors that
determine its index in the real problem are accompanied by $N$ purely imaginary
Takagi vectors, pointing into the complex plane and each with the negative
eigenvalue of its partner. A real minimum on the real manifold therefore has
$N$ downward directions alongside its $N$ upward ones, all pointing directly
into complex configuration space.

The effect of changing the argument of $\beta$ is revealed by
\eqref{eq:generalized.eigenproblem}. Writing $\beta=|\beta|e^{i\phi}$ and
dividing both sides by $|\beta|e^{i\phi/2}$, one finds
\begin{equation}
  \operatorname{Hess}\mathcal S(e^{i\phi/2}v)
  =\frac{\lambda}{|\beta|}e^{-i\phi/2}v^*
  =\frac{\lambda}{|\beta|}(e^{i\phi/2}v)^*
\end{equation}
Therefore, one only needs to consider solutions to the Takagi problem for the
action alone, $\operatorname{Hess}\mathcal Sv_0=\lambda_0 v_0^*$, and then rotate the
resulting Takagi vectors by a constant phase corresponding to half the argument of
$\beta$, or $v(\phi)=v_0e^{-i\phi/2}$. One can see this in the examples of Figs. \ref{fig:1d.stokes} and
\ref{fig:thimble.orientation}, where increasing the argument of $\beta$ from
left to right produces a clockwise rotation of the thimbles in the
complex-$\theta$ plane.

The eigenvalues associated with the Takagi vectors can be further related to properties of the
complex symmetric matrix $\beta\operatorname{Hess}\mathcal S$. Suppose that
$u\in\mathbb R^N$ satisfies the eigenvalue equation
\begin{equation}
  (\beta\operatorname{Hess}\mathcal S)^\dagger(\beta\operatorname{Hess}\mathcal S)u
  =\sigma u
\end{equation}
for some positive real $\sigma$ (because $(\beta\operatorname{Hess}
S)^\dagger(\beta\operatorname{Hess}\mathcal S)$ is self-adjoint and positive definite). The square root of these
numbers, $\sqrt{\sigma}$, are the definition of the \emph{singular values} of
$\beta\operatorname{Hess}\mathcal S$. A direct relationship between these singular
values and the eigenvalues of the real hessian immediately follows by taking a
Takagi vector $v\in\mathbb C$ that satisfies \eref{eq:generalized.eigenproblem},
and writing
\begin{equation}
  \eqalign{
    \sigma v^\dagger u
    &=v^\dagger(\beta\operatorname{Hess}\mathcal S)^\dagger(\beta\operatorname{Hess}\mathcal S)u
    =(\beta\operatorname{Hess}\mathcal Sv)^\dagger(\beta\operatorname{Hess}\mathcal S)u\\
    &=(\lambda v^*)^\dagger(\beta\operatorname{Hess}\mathcal S)u
    =\lambda v^T(\beta\operatorname{Hess}\mathcal S)u
    =\lambda^2 v^\dagger u
  }
\end{equation}
Thus if $v^\dagger u\neq0$, $\lambda^2=\sigma$. It follows that the eigenvalues
of the real hessian are the singular values of the complex matrix
$\beta\operatorname{Hess}\mathcal S$, and the Takagi vectors coincide with the
eigenvectors of the singular value problem up to a complex factor.

\subsection{Evaluating thimble integrals}
\label{subsec:thimble.evaluation}

After all the work of decomposing an integral into a sum over thimbles, one
eventually wants to evaluate it. For large $|\beta|$ and in the
absence of any Stokes points, one can come to a nice asymptotic expression. For a
thorough account of evaluating these integrals (including \emph{at} Stokes
points), see Howls \cite{Howls_1997_Hyperasymptotics}.

Suppose that $\sigma\in\Sigma$ is a stationary point at $s_\sigma\in\tilde\Omega$
with a thimble $\mathcal J_\sigma$ that is not involved in any upstream Stokes points.
Define its contribution to the partition function (neglecting the integer
weight) as
\begin{equation}
  Z_\sigma=\oint_{\mathcal J_\sigma}ds\,e^{-\beta\mathcal S(s)}
\end{equation}
To evaluate this contour integral in the limit of large $|\beta|$, we will make
use of the saddle point method, since the integral will be dominated by its
value at and around the stationary point, where the real part of the action is by
construction at its minimum on the thimble and the integrand is therefore
largest.

We will make a change of coordinates $u(s):\mathcal J_\sigma\to\mathbb R^D$, where $D$ is the dimension of the manifold ($D=N-1$ for the spherical models),  such that
\begin{equation} \label{eq:thimble.integration.def}
  \beta\mathcal S(s)=\beta\mathcal S(s_\sigma)+\frac{|\beta|}2 u(s)^Tu(s)
\end{equation}
\emph{and} the direction of each $\partial u/\partial s$ is aligned with the direction
of the contour. This is possible because, in the absence of any Stokes points,
the eigenvectors of the hessian at the stationary point associated with positive
eigenvalues provide a basis for the thimble. The coordinates $u$ can be real
because the imaginary part of the action is constant on the thimble, and
therefore stays with the value it holds at the stationary point, and the real
part is at its minimum. The preimage of $u(s)^Tu(s)$ gives the vanishing cycles
of the thimble, discussed in an earlier subsection.

The coordinates $u$ can be constructed implicitly in the close vicinity of the stationary point, with their inverse being
\begin{equation}
  s(u)=s_\sigma+\sum_{i=1}^{D}\sqrt{\frac{|\beta|}{\lambda^{(i)}}}v^{(i)}u_i+O(u^2)
\end{equation}
where the sum is over pairs $(\lambda, v)$ which satisfy
\eqref{eq:generalized.eigenproblem} and have $\lambda>0$. It is straightforward
to confirm that these coordinates satisfy \eqref{eq:thimble.integration.def} asymptotically close to the stationary point, as
\begin{equation}
  \eqalign{
    \beta\mathcal S(s(u))
    &=\beta\mathcal S(s_\sigma)
    +\frac12(s(u)-s_\sigma)^T(\beta\operatorname{Hess}\mathcal S)(s(u)-s_\sigma)+\cdots \\
    &=\beta\mathcal S(s_\sigma)
    +\frac{|\beta|}2\sum_{ij}\frac{v^{(i)}_k}{\sqrt{\lambda^{(i)}}}(\beta[\operatorname{Hess}\mathcal S]_{k\ell})\frac{v^{(j)}_\ell}{\sqrt{\lambda^{(j)}}}u_iu_j+\cdots \\
    &=\beta\mathcal S(s_\sigma)
    +\frac{|\beta|}2\sum_{ij}\frac{v^{(i)}_k}{\sqrt{\lambda^{(i)}}}\frac{\lambda^{(j)}(v^{(j)}_k)^*}{\sqrt{\lambda^{(j)}}}u_iu_j+\cdots \\
    &=\beta\mathcal S(s_\sigma)
    +\frac{|\beta|}2\sum_{ij}\frac{\sqrt{\lambda^{(j)}}}{\sqrt{\lambda^{(i)}}}\delta_{ij}u_iu_j+\cdots \\
    &=\beta\mathcal S(s_\sigma)
    +\frac{|\beta|}2\sum_iu_i^2+\cdots
  }
\end{equation}
The Jacobian of this transformation is
\begin{equation}
  \frac{\partial s_i}{\partial u_j}=\sqrt{\frac{|\beta|}{\lambda^{(j)}}}v^{(j)}_i+\cdots
  =\sqrt{\frac1{\lambda_0^{(j)}}}v^{(j)}_i+\cdots
\end{equation}
where $\lambda_0^{(j)}=\lambda^{(j)}/|\beta|$ is the $j$th eigenvalue of the
hessian evaluated at the stationary point for $\beta=1$. This is na\"ively an
$N\times D$ matrix, because the Takagi vectors are $N$ dimensional, but care
must be taken to project each into the tangent space of the manifold to produce
a $D\times D$ matrix. This lets us write $U_{ij}=v_i^{(j)}$ a $D\times D$
unitary matrix, whose determinant will give the correct phase for the measure.

We therefore have
\begin{equation}
Z_\sigma=e^{-\beta\mathcal S(s_\sigma)}\int du\,\det\frac{ds}{du}e^{-\frac{|\beta|}2u^Tu}
\end{equation}
which is exact.
Now we take the saddle point approximation, assuming the integral is dominated
by its value at the stationary point,  and therefore that the determinant can be
approximated by its value at the stationary point. This gives
\begin{equation}
  \eqalign{
    Z_\sigma
      &\simeq e^{-\beta\mathcal S(s_\sigma)}\left.\det\frac{ds}{du}\right|_{s=s_\sigma}\int du\,e^{-\frac{|\beta|}2u^Tu} \\
      &=e^{-\beta\mathcal S(s_\sigma)}\left(\prod_i^D\sqrt{\frac1{\lambda_0^{(i)}}}\right)\det U\left(\frac{2\pi}{|\beta|}\right)^{D/2} \\
      &=e^{-\beta\mathcal S(s_\sigma)}|\det\operatorname{Hess}\mathcal S(s_\sigma)|^{-1/2}\det U\left(\frac{2\pi}{|\beta|}\right)^{D/2}
  }
\end{equation}
We are left with evaluating the determinant of the unitary part of the coordinate transformation.
In circumstances you may be used to, only the absolute value of the determinant
from the coordinate transformation is relevant, and since the determinant of a
unitary matrix is always magnitude one, it doesn't enter the computation.
However, because we are dealing with a contour integral, the directions matter,
and there is not an absolute value around the determinant. Therefore, we must
determine the phase that it contributes.

This is difficult in general, but for real stationary points it can be reasoned
out easily. Take the convention that direction of
contours along the real line is with the standard orientation. Then, when $\beta=1$ a
stationary point of index $k$ has $D-k$ real Takagi vectors and $k$ purely
imaginary Takagi vectors that correspond with upward directions in the flow
and contribute to its thimble. The matrix of Takagi vectors can therefore be
written $U=i^kO$ for an orthogonal matrix $O$, and with all eigenvectors
canonically oriented $\det O=1$. We therefore have $\det U=i^k$ when $\beta=1$. As the
argument of $\beta$ is changed, we know how the eigenvectors change: by a
factor of $e^{-i\phi/2}$ for $\phi=\arg\beta$. Therefore, the contribution for general $\beta$ is
$\det U=(e^{-i\phi/2})^Di^k$. We therefore have, for real stationary points of a real
action,
\begin{equation} \label{eq:real.thimble.partition.function}
  \eqalign{
    Z_\sigma&\simeq\left(\frac{2\pi}{|\beta|}\right)^{D/2}e^{-i\phi D/2}i^{k_\sigma}|\det\operatorname{Hess}\mathcal S(s_\sigma)|^{-\frac12}e^{-\beta\mathcal S(s_\sigma)} \\
                   &=\left(\frac{2\pi}\beta\right)^{D/2}i^{k_\sigma}|\det\operatorname{Hess}\mathcal S(s_\sigma)|^{-\frac12}e^{-\beta\mathcal S(s_\sigma)} \\
  }
\end{equation}

We can see that the large-$\beta$ approximation is consistent with the
relationship between thimble orientation and integer weight outlined in
Fig.~\ref{fig:thimble.orientation}. There, it is seen that taking the argument
of $\beta$ through zero results in a series of Stokes points among real
stationary points of a real action which switches the sign of the integer
weights of thimbles with odd index and preserves the integer weights of
thimbles with even index. For an real action, taking $\beta\to\beta^*$ should
simply take $Z\to Z^*$. Using the formula above, we find
\begin{equation}
  Z(\beta)^*
  =\sum_{\sigma\in\Sigma_0}n_\sigma Z_\sigma(\beta)^*
  =\sum_{\sigma\in\Sigma_0}n_\sigma(-1)^{k_\sigma}Z_\sigma(\beta^*)
  =Z(\beta^*)
\end{equation}
as expected.

\section{The ensemble of symmetric complex-normal matrices}

Having introduced the general method for analytic continuation, we will now
begin dealing with the implications of actions defined in many dimensions
with disorder.  We saw in \S\ref{subsec:stationary.hessian} that the singular
values of the complex hessian of the action at each stationary point are
important to the study of thimbles. Hessians are symmetric matrices by
construction. For real actions of real variables, the study of random symmetric
matrices with Gaussian entries provides insight into a wide variety of
problems. In our case, we will find the relevant ensemble is that of random
symmetric matrices with \emph{complex-normal} entries. In this section, we will
introduce this distribution, review its known properties, and derive its
singular value distribution in the large-matrix limit.

The complex normal distribution with zero mean is the unique Gaussian
distribution in one complex variable $Z$ whose variances are
$\overline{Z^*Z}=\overline{|Z|^2}=\Gamma$ and $\overline{Z^2}=C$. $\Gamma$ is
positive, and $|C|\leq\Gamma$. The special case of $C=\Gamma$, where the
variance of the complex variable and its covariance with its conjugate are the
same, reduces to the ordinary normal distribution. The case where $C=0$ results
in the real and imaginary parts of $Z$ being uncorrelated, in what is known as
the standard complex normal distribution. The probability density function for general $\Gamma$ and $C$ is
defined by
\begin{equation}
  p(z\mid\Gamma,C)=
  \frac1{\pi\sqrt{\Gamma^2-|C|^2}}\exp\left\{
    \frac12\left[\matrix{z^*&z}\right]\left[\matrix{
        \Gamma & C \cr C^* & \Gamma
    }\right]^{-1}\left[\matrix{z\cr z^*}\right]
  \right\}
\end{equation}
This is the same as writing $Z=X+iY$ and requiring that the mutual distribution
in $X$ and $Y$ be normal with $\overline{X^2}=\Gamma+\operatorname{Re}C$,
$\overline{Y^2}=\Gamma-\operatorname{Re}C$, and
$\overline{XY}=\operatorname{Im}C$.

\begin{figure}
  \centering

  \includegraphics{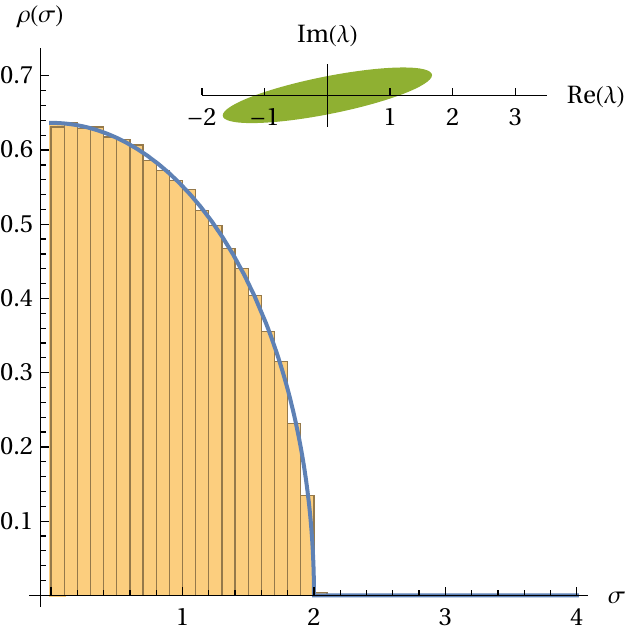}
  \includegraphics{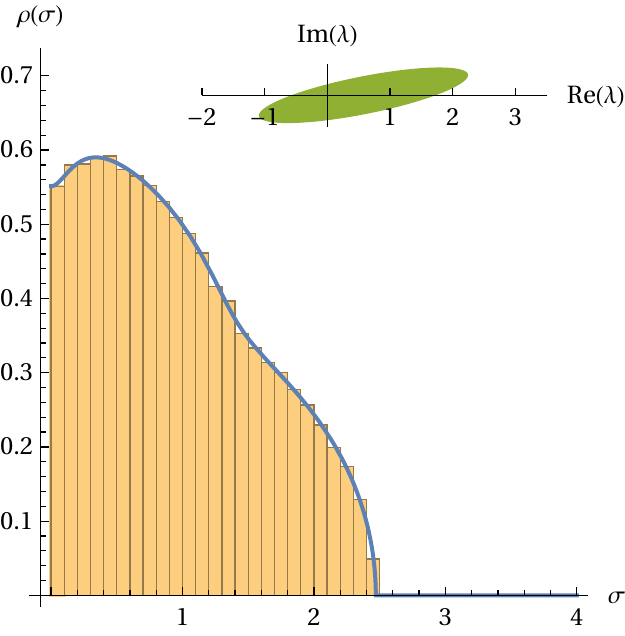}\\
  \includegraphics{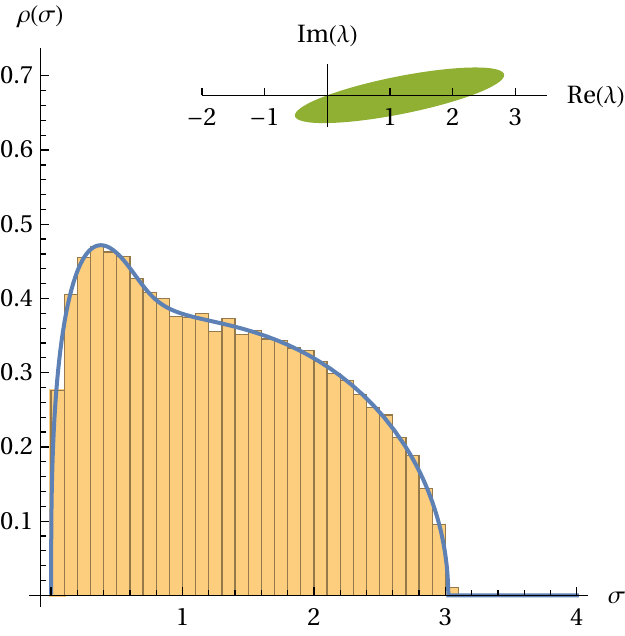}
  \includegraphics{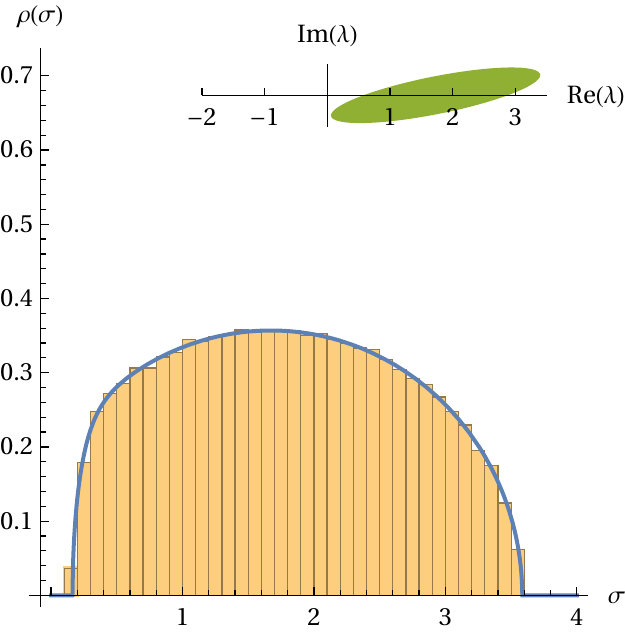}

  \caption{
    Eigenvalue and singular value spectra of a random matrix $B=A+\lambda_0I$, where the entries of $A$ are complex-normal distributed with $N\overline{|A_{ij}|^2}=\Gamma_0=1$ and $N\overline{A_{ij}^2}=C_0=\frac7{10}e^{i\pi/8}$.
    The diagonal shifts differ in each plot, with (a) $\lambda_0=0$, (b)
    $\lambda_0=\frac12|\lambda_{\mathrm{gap}}|$, (c)
    $\lambda_0=|\lambda_{\mathrm{gap}}|$, and (d)
    $\lambda_0=\frac32|\lambda_{\mathrm{gap}}|$. The shaded region of each
    inset shows the support of the eigenvalue distribution \eqref{eq:ellipse}.
    The solid line on each plot shows the distribution of singular values
    \eqref{eq:spectral.density}, while the overlaid histogram shows the
    empirical distribution from $2^{10}\times2^{10}$ complex normal matrices.
  } \label{fig:spectra}
\end{figure}

We will consider an ensemble of random $N\times N$ matrices $B=A+\lambda_0I$, where the
entries of $A$ are complex-normal distributed with variances
$\overline{|A_{ij}|^2}=\Gamma_0/N$ and $\overline{A_{ij}^2}=C_0/N$, and $\lambda_0$ is a
constant shift to the  diagonal. The eigenvalue distribution of the matrices $A$
is already known to take the form of an elliptical ensemble in the large-$N$
limit, with constant support inside the ellipse defined by
\begin{equation} \label{eq:ellipse}
  \left(\frac{\operatorname{Re}(\lambda e^{i\theta})}{1+|C_0|/\Gamma_0}\right)^2+
  \left(\frac{\operatorname{Im}(\lambda e^{i\theta})}{1-|C_0|/\Gamma_0}\right)^2
  <\Gamma_0
\end{equation}
where $\theta=\frac12\arg C_0$ \cite{Nguyen_2014_The}. The eigenvalue
spectrum of $B$ is therefore constant inside the same ellipse
translated so that its center lies at $\lambda_0$.  Examples of these
distributions are shown in the insets of Fig.~\ref{fig:spectra}.

When $C=0$ and the elements of $A$ are standard complex normal, the singular
value distribution of $B$ is a complex Wishart distribution. For $C\neq0$ the
problem changes, and to our knowledge a closed form of the singular value distribution is not in the literature.
We have worked out an implicit form for the singular value spectrum using the
replica method, first published in \cite{Kent-Dobias_2021_Complex}.

The singular values of $B$ correspond with the square-root of the eigenvalues
of $B^\dagger B$, but also they correspond to the absolute value of the
eigenvalues of the real $2N\times2N$ block matrix
\begin{equation}
  \left[\matrix{\operatorname{Re}B&-\operatorname{Im}B\cr-\operatorname{Im}B&-\operatorname{Re}B}\right]
\end{equation}
as we saw in \S\ref{subsec:stationary.hessian}. The $2N\times2N$ problem is easier
to treat analytically than the $N\times N$ one because the matrix under study
is linear in the entries of $B$. The eigenvalue spectrum of this block matrix
can be studied by ordinary techniques from random matrix theory.  Defining the `partition function'
\begin{equation} \fl \qquad
  Z(\sigma)=\int dx\,dy\,\exp\left\{
    -\frac12\left[\matrix{x&y}\right]
    \left(\sigma I-
    \left[\matrix{\hphantom{-}\operatorname{Re}B&-\operatorname{Im}B\cr-\operatorname{Im}B&-\operatorname{Re}B}\right]
  \right)
    \left[\matrix{x\cr y}\right]
  \right\}
\end{equation}
implies a Green function
\begin{equation}
  G(\sigma)=\frac\partial{\partial\sigma}\log Z(\sigma)
\end{equation}
whose poles give the singular values of $B$.
This can be put into a manifestly complex form using the method of
\S\ref{subsec:stationary.hessian}, with the same linear transformation of
$x,y\in\mathbb R^N$ into $z\in\mathbb C^N$. This gives
\begin{equation}
  \eqalign{
    Z(\sigma)
    &=\int dz^*dz\,\exp\left\{
      -\frac12\left[\matrix{z^*&-iz}\right]
      \left(\sigma I-
      \left[\matrix{0&(iB)^*\cr iB&0}\right]
    \right)
      \left[\matrix{z\cr iz^*}\right]
    \right\} \\
    &=\int dz^*dz\,\exp\left\{
      -\frac12\left(
        2z^\dagger z\sigma-z^\dagger B^*z^*-z^TBz
      \right)
    \right\} \\
    &=\int dz^*dz\,\exp\left\{
      -z^\dagger z\sigma+\operatorname{Re}(z^TBz)
    \right\}
  }
\end{equation}
which is a general expression for the singular values $\sigma$ of a symmetric
complex matrix $B$.

Introducing replicas to eliminate the logarithm in the
Green function \cite{Livan_2018_Introduction} gives
\begin{equation} \label{eq:green.replicas} \fl\quad
  G(\sigma)=\lim_{n\to0}\int dz^*dz\,z_0^\dagger z_0
    \exp\left\{
    -\sum_\alpha^n\left[z_\alpha^\dagger z_\alpha\sigma
      +\operatorname{Re}\left(z_\alpha^TBz_\alpha\right)
    \right]
  \right\}
\end{equation}
The average is then made over
the entries of $B$ and Hubbard--Stratonovich is used to change variables to the
replica matrices
$N\alpha_{\alpha\beta}=z_\alpha^\dagger z_\beta$ and
$N\chi_{\alpha\beta}=z_\alpha^Tz_\beta$, and a series of
replica vectors. The replica-symmetric ansatz leaves all replica vectors
zero, and $\alpha_{\alpha\beta}=\alpha_0\delta_{\alpha\beta}$,
$\chi_{\alpha\beta}=\chi_0\delta_{\alpha\beta}$. The result is
\begin{equation}\label{eq:green.saddle}
  \eqalign{
    \overline G(\sigma)=N\lim_{n\to0}\int &d\alpha_0\,d\chi_0^*\,d\chi_0\,\alpha_0
    \exp\left\{nN\left[
      1+\frac18\Gamma_0\alpha_0^2-\frac{\alpha_0\sigma}2\right.\right.\cr
         &\left.\left.+\frac12\log(\alpha_0^2-|\chi_0|^2)+\frac12\operatorname{Re}\left(\frac14C_0^*\chi_0^2+\lambda_0^*\chi_0\right)
    \right]\right\}.
  }
\end{equation}

The argument of the exponential has several saddles. The solutions $\alpha_0$
are the roots of a sixth-order polynomial, and the root with the smallest value
of $\operatorname{Re}\alpha_0$ gives the correct solution in all the cases we
studied. A detailed analysis of the saddle point integration is needed to
understand why this is so. Evaluated at such a solution, the density of
singular values follows from the jump across the cut in the infinite-$N$ limit, or
\begin{equation} \label{eq:spectral.density}
  \rho(\sigma)=\frac1{i\pi N}\left(
    \lim_{\operatorname{Im}\sigma\to0^+}\overline G(\sigma)
    -\lim_{\operatorname{Im}\sigma\to0^-}\overline G(\sigma)
  \right)
\end{equation}
Examples of this distribution can be seen in Fig.~\ref{fig:spectra} compared with numeric
experiments.

The formation of a gap in the singular value spectrum naturally corresponds to
the origin leaving the support of the eigenvalue spectrum.  Weyl's theorem
requires that the product over the norm of all eigenvalues must not be greater
than the product over all singular values \cite{Weyl_1912_Das}.  Therefore, the
absence of zero eigenvalues implies the absence of zero singular values. The
determination of the constant shift $\lambda_0$ at which the distribution
of singular values becomes gapped is reduced to the geometry problem of
determining when the boundary of the ellipse defined in \eqref{eq:ellipse}
intersects the origin, and yields
\begin{equation} \label{eq:gap.eigenvalue}
  |\lambda_{\mathrm{gap}}|^2
  =\Gamma_0\frac{(1-|\delta|^2)^2}
  {1+|\delta|^2-2|\delta|\cos(\arg\delta+2\arg\lambda_0)}
\end{equation}
for $\delta=C_0/\Gamma_0$. Because the support is an ellipse, this naturally
depends on the argument of $\lambda_0$, or the direction in the complex plane
in which the distribution is shifted.

\section{The \textit{p}-spin spherical models}
\label{sec:p-spin}

The $p$-spin spherical models are defined by the action
\begin{equation} \label{eq:p-spin.hamiltonian}
  \mathcal S(x)=\sum_{p=2}^\infty a_p\mathcal S_p(x)
\end{equation}
which is a sum of the `pure' $p$-spin actions
\begin{equation} \label{eq:pure.p-spin.hamiltonian}
  \mathcal S_p(x)=\frac1{p!}\sum_{i_1\cdots i_p}J_{i_1\cdots i_p}x_{i_1}\cdots x_{i_p}
\end{equation}
The variables $x\in\mathbb R^N$ are constrained to lie on the sphere $x^2=N$,
making the model $D=N-1$ dimensional. The couplings $J$ form totally symmetric
$p$-tensors whose components are normally distributed with zero mean and
variance $\overline{J^2}=p!/2N^{p-1}$. The `pure' $p$-spin models have
$a_i=\delta_{ip}$, while the mixed have some more complicated set of coefficients $a$.

The configuration space manifold $\Omega=\{x\mid x^2=N, x\in\mathbb R^N\}$ has a
complex extension $\tilde\Omega=\{z\mid z^2=N, z\in\mathbb C^N\}$. The natural
extension of the Hamiltonian \eref{eq:p-spin.hamiltonian} to this complex
manifold by replacing $x$ with $z\in\mathbb C^N$ is holomorphic. The normal to
this manifold at any point $z\in\tilde\Omega$ is always in the direction $z$.
The projection operator onto the tangent space of this manifold is given by
\begin{equation}
  P=I-\frac{zz^\dagger}{|z|^2},
\end{equation}
where indeed $Pz=z-z|z|^2/|z|^2=0$ and $Pz'=z'$ for any $z'$ orthogonal to $z$.
When studying stationary points, the constraint can be added to the action
using a Lagrange multiplier $\mu$ by writing
\begin{equation}
  \tilde\mathcal S(z)=\mathcal S(z)-\frac\mu2(z^Tz-N)
\end{equation}
The gradient of the constraint is simple with $\partial g=z$, and \eqref{eq:multiplier} implies that
\begin{equation}
  \mu
  =\frac1Nz^T\partial\mathcal S
  =\sum_{p=2}^\infty a_pp\frac{\mathcal S_p(z)}N
\end{equation}
For the pure $p$-spin in particular this implies that $\mu=p\epsilon$ for
specific energy $\epsilon=\mathcal S_p/N$.

\subsection{2-spin}
\label{subsec:2-spin}

The pure 2-spin model is diagonalizable and therefore exactly solvable, and is
not complex in the sense of having a superextensive number of stationary points
in its action. However, it makes a good exercise of how the ideas of analytic
continuation will apply in the literally more complex case of the $p$-spin for $p>2$.
The Hamiltonian of the pure $2$-spin model is defined by
\begin{equation}
  \mathcal S_2(z)=\frac12z^TJz.
\end{equation}
where the matrix $J$ is generically diagonalizable. In a diagonal basis,
$J_{ij}=\lambda_i\delta_{ij}$. Then $\partial_i H=\lambda_iz_i$. We will
henceforth assume to be working in this basis. The constrained action is
\begin{equation}
  \tilde\mathcal S(z)=\mathcal S_2(z)-\epsilon(z^Tz-N)
\end{equation}
Stationary points must satisfy
\begin{equation}
  0=\partial_i\tilde\mathcal S=(\lambda_i-2\epsilon)z_i
\end{equation}
which is only possible for $z_i=0$ or $\epsilon=\frac12\lambda_i$. Generically
the $\lambda_i$ will all differ, so this can only be satisfied for one
$\lambda_i$ at a time, and to be a stationary point all other $z_j$ must be
zero. In the direction in question,
\begin{equation}
  \frac1N\frac12\lambda_iz_i^2=\epsilon=\frac12\lambda_i,
\end{equation}
whence $z_i=\pm\sqrt{N}$. Thus there are $2N$ stationary points, each
corresponding to $\pm$ the cardinal directions on the sphere in the diagonalized basis. The
energy at each stationary point is real if the couplings are real, and
therefore there are no complex stationary points in the ordinary 2-spin model.

\begin{figure}
  \hfill\includegraphics{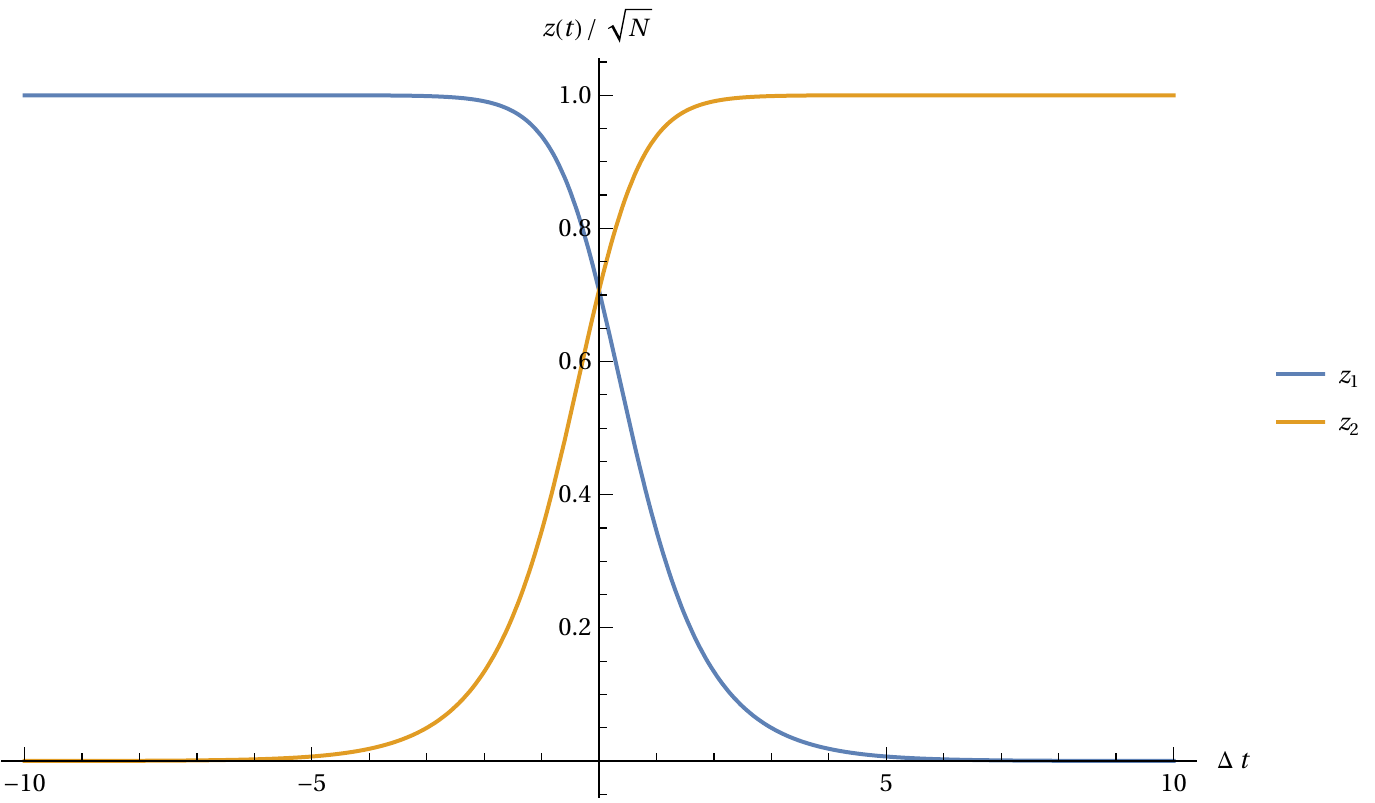}

  \caption{
    The Stokes line in the 2-spin model when the stationary points associated
    with the first and second cardinal directions are brought to the same
    imaginary energy. $\Delta$ is proportional to the difference between the
    real energies of the first and the second stationary point; when $\Delta >0$
    flow is from first to second, while when $\Delta < 0$ it is reversed.
  } \label{fig:two-spin}
\end{figure}

Imagine for a moment that the coupling are allowed to be complex, giving the
stationary points of the model complex energies and therefore potentially
interesting thimble structure. Generically, the eigenvalues of the coupling
matrix will have distinct imaginary parts, and there will be no Stokes lines.
Suppose that two stationary points are brought to the same imaginary energy by
some continuation; without loss of generality, assume these are associated with
the first and second cardinal directions. Since the gradient is proportional to
$z$, any components that are zero at some time will be zero at all times. The
gradient flow dynamics for the two components of interest assuming all others are
zero are
\begin{equation}
  \dot z_1
  =-z_1^*\left(\lambda_1^*-\frac{\lambda_1^*z_1^*z_1+\lambda_2^*z_2^*z_2}{|z_1|^2+|z_2|^2}\right)
  =-(\lambda_1-\lambda_2)^*z_1^*\frac{|z_2|^2}{|z_1|^2+|z_2|^2}
\end{equation}
and the same for $z_2$ with all indices swapped.  Since $\Delta=\lambda_1-\lambda_2$ is
real when the energies and therefore eigenvalues have the same imaginary part, if $z_1$ begins real it remains real, with the same for $z_2$. Since the
stationary points are at real $z$, we make this restriction, and find
\begin{equation}
  \frac d{dt}(z_1^2+z_2^2)=0 \qquad
  \frac d{dt}\frac{z_2}{z_1}=\Delta\frac{z_2}{z_1}
\end{equation}
Therefore $z_2/z_1=e^{\Delta t}$, with $z_1^2+z_2^2=N$ as necessary.  Depending
on the sign of $\Delta$, $z$ flows from one stationary point to the other over
infinite time. This is a Stokes line, and establishes that any two distinct stationary
points in the 2-spin model with the same imaginary energy will possess one.
These trajectories are plotted in Fig.~\ref{fig:two-spin}.

Since they sit at the corners of a simplex, the distinct stationary points of the 2-spin
model are all adjacent: no stationary point is separated from another by the
separatrix of a third. This means that when the imaginary energies of two
stationary points are brought to the same value, their surfaces of constant
imaginary energy join. However, this is not true for stationary points related
by the symmetry $z\to-z$, as seen in Fig.~\ref{fig:3d.thimbles}.

Since the 2-spin model with real couplings does not have any stationary points
in the complex plane, analytic continuation can be made without any fear of
running into Stokes points. Starting from real, large $\beta$, making an
infinitesimal phase rotation into the complex plane results in a decomposition
into thimbles where that of each stationary point is necessary, because all
stationary points are real and their antithimbles all intersect the real sphere. The curvature of the action at the stationary
point lying at $z_i=\sqrt N\delta_{ik}$ in the $j$th direction is given by
$\lambda_k-\lambda_j=2(\epsilon_k-\epsilon_j)$. Therefore the generic case of
$N$ distinct eigenvalues of the coupling matrix leads to $2N$ stationary points
with $N$ distinct energies, two at each index from $0$ to $D=N-1$. Starting with
the expression \eqref{eq:real.thimble.partition.function}, we
have
\begin{equation}
  \eqalign{
    Z
    &=\int_{S^{N-1}}ds\,e^{-\beta\mathcal S_2(s)}
    =\sum_{\sigma\in\Sigma_0}n_\sigma\int_{\mathcal J_\sigma}ds\,e^{-\beta\mathcal S_2(s)} \\
    &\simeq\sum_{\sigma\in\Sigma_0}i^{k_\sigma}\left(\frac{2\pi}\beta\right)^{D/2}e^{-\beta\mathcal S_2(s_\sigma)}|\det\operatorname{Hess}\mathcal S_2(s_\sigma)|^{-\frac12} \\
    &=2\sum_{k=0}^D\exp\left\{
      i\frac\pi2k+\frac D2\log\frac{2\pi}\beta-N\beta\epsilon_k-\frac12\sum_{\ell\neq k}\log2|\epsilon_k-\epsilon_\ell|
    \right\}
  }
\end{equation}
where $\epsilon_k$ is the energy of the twin stationary points of index $k$. In the large $N$ limit, we take advantage of the limiting distribution $\rho$ of these energies to write
\begin{equation} \fl
  \eqalign{
    \overline Z
    &\simeq2\int d\epsilon\,\rho(\epsilon)\exp\left\{
      i\frac\pi2k_\epsilon+\frac D2\log\frac{2\pi}\beta-N\beta\epsilon-\frac D2\int d\epsilon'\,\rho(\epsilon')\log2|\epsilon-\epsilon'|
    \right\} \\
    &=2\int d\epsilon\,\rho(\epsilon)e^{Nf(\epsilon)}
  }
\end{equation}
Since the $J$ of the 2-spin model is a symmetric real matrix with variance
$1/N$, its eigenvalues are distributed by a semicircle distribution of radius 2,
and therefore the energies $\epsilon$ are distributed by a semicircle
distribution of radius one, with
\begin{equation}
  \rho(\epsilon)=\frac2{\pi}\sqrt{1-\epsilon^2}
\end{equation}
The index as a function of energy level is given by the cumulative density function
\begin{equation}
  k_\epsilon=D\int_{-\infty}^\epsilon d\epsilon'\,\rho(\epsilon')=\frac D\pi\left(
    \epsilon\sqrt{1^2-\epsilon^2}+2\tan^{-1}\frac{1+\epsilon}{\sqrt{1-\epsilon^2}}
  \right)
\end{equation}
Finally, the product over the singular values corresponding to descending directions gives
\begin{equation}
  \frac12\int d\epsilon'\,\rho(\epsilon')\log2|\epsilon-\epsilon'|
  =-\frac14+\frac12\epsilon^2
\end{equation}
for $\epsilon^2<1$. This gives the function $f$ in the exponential as
\begin{equation}
  \operatorname{Re}f=-\epsilon\operatorname{Re}\beta+\frac14-\frac12\epsilon^2
\end{equation}
\begin{equation}
  \operatorname{Im}f=-\epsilon\operatorname{Im}\beta+\frac12\left(
    \epsilon\sqrt{1-\epsilon^2}+2\tan^{-1}\frac{1+\epsilon}{\sqrt{1-\epsilon^2}}
  \right)
\end{equation}
The value of the integral will be dominated by the contribution near the
maximum of the real part of $f$, which is
\begin{equation}
  \epsilon_{\mathrm{max}}=\left\{\matrix{-\operatorname{Re}\beta & \operatorname{Re}\beta\leq1\cr -1 & \mathrm{otherwise}}\right.
\end{equation}
For $\operatorname{Re}\beta>1$, the maximum is concentrated in the ground state
and the real part of $f$ comes to a cusp, meaning that the oscillations do not
interfere in taking the saddle point. Once this line is crossed and the maximum
enters the bulk of the spectrum, one expects to find cancellations caused by
the incoherent contributions of thimbles with nearby energies to
$\epsilon_{\mathrm{max}}$. Therefore, one expects that $\overline Z$ enters a
phase with no coherent average when $\operatorname{Re}\beta=1$.

On the other hand, there is another point where the thimble sum becomes
coherent. This is when the oscillation frequency near the maximum energy goes
to zero. This happens for
\begin{equation}
  0
  =\frac{\partial}{\partial\epsilon}\operatorname{Im}f\Big|_{\epsilon=\epsilon_{\mathrm{max}}}
  =-\operatorname{Im}\beta+\sqrt{1-\epsilon_{\mathrm{max}}^2}
  =-\operatorname{Im}\beta+\sqrt{1-(\operatorname{Re}\beta)^2}
\end{equation}
or for $|\beta|=1$. Here the sum of contributions from thimbles near the
maximum again becomes coherent, because the period of oscillations in
$\epsilon$ diverges at the maximum. These conditions correspond precisely to
the phase boundaries of the density of zeros in the 2-spin model found previously using other methods
\cite{Obuchi_2012_Partition-function, Takahashi_2013_Zeros}.

We've seen that even in the 2-spin model, which is not complex, making a
thimble decomposition in a theory with many saddles does not necessarily fix
the sign problem. Instead, it takes a potentially high-dimensional sign problem
and produces a one-dimensional one, represented by the oscillatory integral
over $e^{Nf(\epsilon)}$. In some regimes, it can be argued that integral has a
maximum with a coherent neighborhood, allowing computation to be made. In
others, oscillations in the phase remain, from the sum over the many thimbles.
We will find a similar story for the pure $p$-spin models for $p>2$ in the next
sections, complicated by the additional presence of Stokes points in the
continuation.

\subsection{Pure \textit{p}-spin: where are the saddles?}
\label{subsec:p-spin.one.replica}

We studied the distribution of stationary points in the pure $p$-spin models in
previous work \cite{Kent-Dobias_2021_Complex}. Here, we will review the
method and elaborate on some of the results relevant to analytic continuation.

The complexity of the real $p$-spin models has been studied extensively, and is even known rigorously \cite{Auffinger_2012_Random}. If $\mathcal N(\epsilon)$ is the number of stationary points with specific energy $\epsilon$, then the complexity is defined by
\begin{equation}
  \Sigma(\epsilon)=\lim_{N\to\infty}\frac1N\log\overline\mathcal N(\epsilon)
\end{equation}
a natural measure of how superextensive the average number $\overline\mathcal
N\sim e^{N\Sigma}$ is. The complexity is also known for saddles of particular
index, with, e.g., $\Sigma_{k=1}$ measuring the complexity of rank-one saddles
and $\Sigma_{k=0}$ measuring that of minima. The minimum energy for which
$\Sigma_{k=0}$ is positive corresponds to the ground state energy of the model,
because at large $N$ below this the number of minima is expected to be
exponentially small with $N$. We'll write the ground state energy as $\epsilon_{k=0}$, and the lowest energies at which rank $j$ saddles are found as $\epsilon_{k=j}$, so that, e.g.,
\begin{equation}
  0=\Sigma(\epsilon_{k=0})=\Sigma_{k=0}(\epsilon_{k=0}) \qquad
  0=\Sigma_{k=1}(\epsilon_{k=1})
\end{equation}

In the real case, the $p$-spin models posses a threshold energy
\begin{equation}
  |\epsilon_{\mathrm{th}}|^2=\frac{2(p-1)}{p}
\end{equation}
below which there are exponentially many minima
compared to saddles, and above which vice versa. This threshold persists in a
more generic form in the complex case, where now the threshold separates
stationary points that have mostly gapped from mostly ungapped spectra. Since
the $p$-spin model has a hessian that consists of a symmetric complex matrix
with a shifted diagonal, we can use the results of
\S\ref{subsec:stationary.hessian}. The variance of the $p$-spin hessian without
shift is
\begin{equation}
  \overline{|\partial\partial\mathcal S_p|^2}
  =\frac{p(p-1)(\frac1Nz^\dagger z)^{p-2}}{2N}
  =\frac{p(p-1)}{2N}(1+2Y)^{p-2}
\end{equation}
\begin{equation}
  \overline{(\partial\partial\mathcal S_p)^2}
  =\frac{p(p-1)(\frac1Nz^Tz)^{p-2}}{2N}
  =\frac{p(p-1)}{2N}
\end{equation}
where $Y=\frac1N\|\operatorname{Im}z\|^2$ is a measure of how far the
stationary point is into the complex configuration space.  As expected for a
real problem, the two variances coincide when $Y=0$.  The diagonal shift is
$-p\epsilon$. In the language of \S\ref{subsec:stationary.hessian}, this means
that $\Gamma_0=p(p-1)(1+2Y)^{p-2}/2$, $C_0=p(p-1)/2$, and
$\lambda_0=-p\epsilon$. This means that the energy at which the gap appears is,
using \eqref{eq:gap.eigenvalue},
\begin{equation}
  |\epsilon_\mathrm{gap}|^2
  =\frac{p-1}{2p}
  \frac{[1-(1+2Y)^{2(p-2)}]^2(1+2Y)^{p-2}}
  {1+(1+2Y)^{2(p-2)}-2(1+2Y)^{p-2}\cos(2\arg\epsilon)}
\end{equation}
When $\epsilon$ is real,
$\lim_{Y\to0}|\epsilon_{\mathrm{gap}}|=|\epsilon_\mathrm{th}|$.

The complexity of stationary points by their energy and location $Y$ can be
determined by the Kac--Rice formula. Any stationary point of the action is a
stationary point of the real part of the action, and we can write
\begin{equation} \label{eq:real.kac-rice}
  \mathcal N
    = \int dx\,dy\,\delta(\partial_x\operatorname{Re}\tilde\mathcal S_p)\delta(\partial_y\operatorname{Re}\tilde\mathcal S_p)
            \left|\det\operatorname{Hess}_{x,y}\operatorname{Re}\mathcal S_p\right|
\end{equation}
This expression is to be averaged over $J$ to give the complexity $\Sigma$ as
$N \Sigma= \overline{\log\mathcal N}$, a calculation that involves the replica
trick. Based on the experience from similar problems
\cite{Castellani_2005_Spin-glass},  the \emph{annealed approximation} $N \Sigma
\sim \log \overline{ \mathcal N}$ is expected to be exact wherever the
complexity is positive.

As in \S\ref{subsec:stationary.hessian}, this expression can be bright into a
manifestly complex form using Cauchy--Riemann relations. This gives
\begin{equation}
  \mathcal N
   =\int dz^*dz\,d\hat z^*d\hat z\,d\eta^*d\eta\,d\gamma^*d\gamma\exp\left\{
    \operatorname{Re}\left(
      \hat z^T\partial\tilde\mathcal S_p+\eta^T\partial\partial\tilde\mathcal S_p\gamma
    \right)
  \right\}
\end{equation}
where $\eta$ and $\gamma$ are $N$-dimensional Grassmann fields. This can be more
conveniently studied using the method of superfields. For an overview of
superfields applied to the $p$-spin spherical models, see
\cite{Kurchan_1992_Supersymmetry}. Our previous work deriving the complexity
does not use superfields \cite{Kent-Dobias_2021_Complex}, but they will be
essential for compactly writing the \emph{two} replica complexity in the next
section, and so we briefly introduce the technique here. Introducing the
one-component Grassmann variables $\theta$ and $\bar\theta$, define the
superfield
\begin{equation}
  \phi(1)=z+\bar\theta(1)\eta+\gamma\theta(1)+\hat z\bar\theta(1)\theta(1)
\end{equation}
and its measure $d\phi=dz\,d\hat z\,d\eta\,d\gamma$.
Then the expression for the number of stationary points can be written in a compact form, as
\begin{equation}
  \mathcal N=\int d\phi^*d\phi\,\exp\left\{
    \int d1\,\operatorname{Re}
      \tilde\mathcal S_p(\phi(1))
  \right\}
\end{equation}
where $d1=d\bar\theta(1)\,d\theta(1)$ denotes the integration over the Grassmann variables.
This can be related to the previous expression by expansion with respect to
the Grassmann variables, recognizing that $\theta^2=\bar\theta^2=0$ restricts
the series to two derivatives.

From here the process can be treated as usual, averaging over the couplings and
replacing bilinear combinations of the fields with their own variables via a
Hubbard--Stratonovich transformation. Defining the supermatrix
\begin{equation}
  Q(1,2)=\frac1N\left[\matrix{
      \phi(1)^T\phi(2)&\phi(1)^T\phi(2)^*\cr
      \phi(1)^\dagger\phi(2)&\phi(1)^\dagger\phi(2)^*
  }\right]
\end{equation}
the result can be written, neglecting constant factors, as an integral over $Q$ like
\begin{equation}
  \overline\mathcal N\simeq\int dQ\,e^{NS_\mathrm{eff}(Q)}
\end{equation}
where the effective action functional $S_\mathrm{eff}$ of the supermatrix $Q$ is
\begin{equation} \fl
  \eqalign{
    S_{\mathrm{eff}}&=
    \int d1\,d2\,\operatorname{Tr}\left(
        \frac14\left[
          \matrix{\frac14&\frac14\cr\frac14&\frac14}
          \right]Q^{(p)}(1,2)-\frac p2\left[
          \matrix{\frac\epsilon2&0\cr0&\frac{\epsilon^*}2}
        \right](Q(1,1)-I)\delta(1,2)
      \right) \\
                                &\hspace{28em}+\frac12\log\det Q
    }
\end{equation}
The exponent in parentheses denotes element-wise exponentiation, and
\begin{equation}
  \delta(1,2)=(\bar\theta(1)-\bar\theta(2))(\theta(1)-\theta(2))
\end{equation}
is the superspace $\delta$-function, and the determinant and trace are a superdeterminant and supertrace, respectively. Algebraically and under calculus they behave nearly like their non-super counterparts.
This leads to the condition for a saddle point of
\begin{equation}
  0
  =\frac{\partial S_\mathrm{eff}}{\partial Q(1,2)}
  =\frac p{16}Q^{(p-1)}(1,2)-\frac p2\left[
    \matrix{\frac\epsilon2&0\cr0&\frac{\epsilon^*}2}
  \right]\delta(1,2)
  +\frac12Q^{-1}(1,2)
\end{equation}
where the inverse supermatrix is defined by
\begin{equation}
  I\delta(1,2)=\int d3\,Q^{-1}(1,3)Q(3,2)
\end{equation}
Convolving both sides by another supermatrix to remove the inverse, we arrive at the saddle point equations
\begin{equation}
  \eqalign{
    0
    &=\int d3\,\frac{\partial S_\mathrm{eff}}{\partial Q(1,3)}Q(3,2) \\
    &=\frac p{16}\int d3\,Q^{(p-1)}(1,3)Q(3,2)-\frac p2\left[
    \matrix{\frac\epsilon2&0\cr0&\frac{\epsilon^*}2}
    \right]Q(1,2)+\frac12I\delta(1,2)
  }
\end{equation}
When expanded, the supermatrix $Q$ contains nine independent bilinear
combinations of the original variables: $z^\dagger z$, $\hat z^T z$, $\hat
z^\dagger z$, $\hat z^T\hat z$, $\hat z^\dagger\hat z$, $\eta^\dagger\eta$,
$\gamma^\dagger\gamma$, $\eta^\dagger\gamma$, and $\eta^T\gamma$. The saddle
point equations can be used to eliminate all but one of these, the `radius'
like term $z^\dagger z$. When combined with the constraint, this term can be
related directly to the magnitude of the imaginary part of $z$, since
$z^\dagger z=x^Tx+y^Ty=N+2y^Ty=N(1+2Y)$ for $Y=\|\operatorname{Im}z\|^2/N=y^Ty/N$. The complexity can then be written in terms of $r=z^\dagger z/N=1+2Y$ as
\begin{equation}
  \Sigma
  =
  \log(p-1)-\frac12\log\left(
    \frac{1-r^{-2(p-1)}}{1-r^{-2}}
  \right)
  -\frac{(\operatorname{Re}\epsilon)^2}{R_+^2}-\frac{(\operatorname{Im}\epsilon)^2}{R_-^2}
  +I_p(\epsilon/|\epsilon_\mathrm{th}|)
\end{equation}
where
\begin{equation}
  R_\pm^2=\frac{p-1}2\frac{(r^{p-2}\pm1)\left[
      r^{2(p-1)}\pm(p-1)r^{p-2}(r^2-1)-1
  \right]}{
    1+r^{2(p-2)}\left[p(p-2)(r^2-1)-1\right]
  }
\end{equation}
and the function $I_p(u)=0$ if $|\epsilon|^2<|\epsilon_\mathrm{gap}|^2$ and
\begin{equation}
  \eqalign{
    I_p(u)
    &=\left(\frac12+\frac1{r^{p-2}-1}\right)^{-1}(\operatorname{Re}u)^2
    -\left(\frac12-\frac1{r^{p-2}+1}\right)^{-1}(\operatorname{Im}u)^2 \\
    &\qquad-\log\left(
      r^{p-2}\left|
      u+\sqrt{u^2-1}
      \right|^2
    \right)+2\operatorname{Re}
    \left(
      u\sqrt{u^2-1}
    \right)
  }
\end{equation}
otherwise. The branch of the square roots are chosen such that the real part of
the root has the opposite sign as the real part of $u$, e.g., if
$\operatorname{Re}u<0$ then $\operatorname{Re}\sqrt{u^2-1}>0$. If the real part
is zero, then the sign is taken so that the imaginary part of the root has the
opposite sign of the imaginary part of $u$.

\begin{figure}
  \hspace{4pc}
  \includegraphics{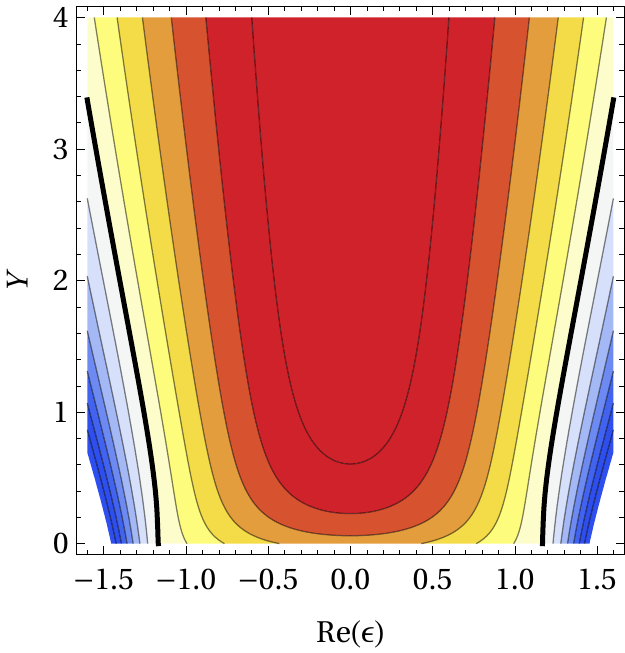}
  \hspace{-2pc}
  \includegraphics{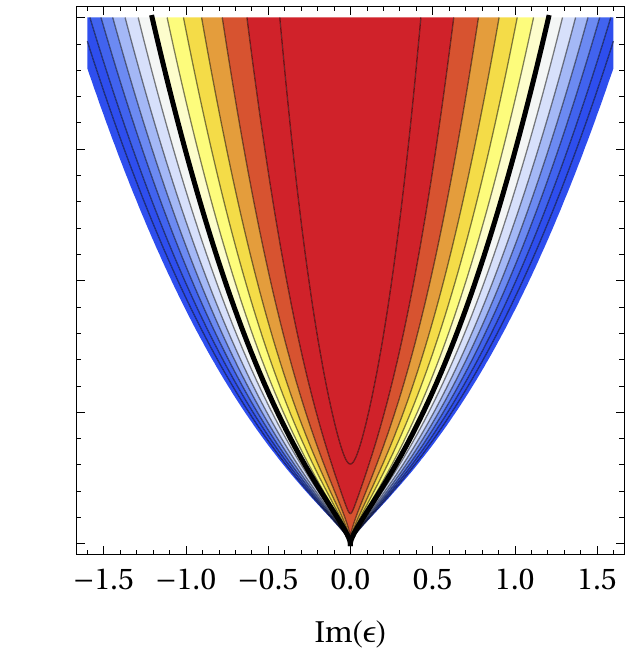}
  \includegraphics{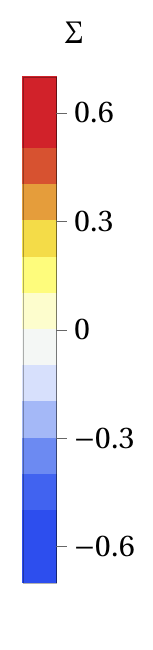}

  \caption{
    The complexity of the 3-spin spherical model in the complex plane, as a
    function of pure real and imaginary energy (left and right) and the
    magnitude $Y=\|\operatorname{Im}z\|^2/N$ of the distance into the complex
    configuration space. The thick black contour shows the line of zero
    complexity, where stationary points become exponentially rare in $N$.
  } \label{fig:p-spin.complexity}
\end{figure}

Contours of this complexity for the pure 3-spin are plotted in
Fig.~\ref{fig:p-spin.complexity} for pure real and imaginary energy. The thick black
line shows the contour of zero complexity, where stationary points are no
longer found at large $N$. As the magnitude of the imaginary part of the spin
taken greater, more stationary points are found, and at a wider array of
energies. This is also true in other directions into the complex energy plane,
where the story is qualitatively the same. At any energy, the limit
$Y\to\infty$ or $r\to\infty$ results in $\Sigma=\log(p-1)$, which saturates the
Bézout bound on the number of stationary points a polynomial of order $p$ can
have \cite{Bezout_1779_Theorie}.

\begin{figure}
  \hspace{2pc}
  \includegraphics{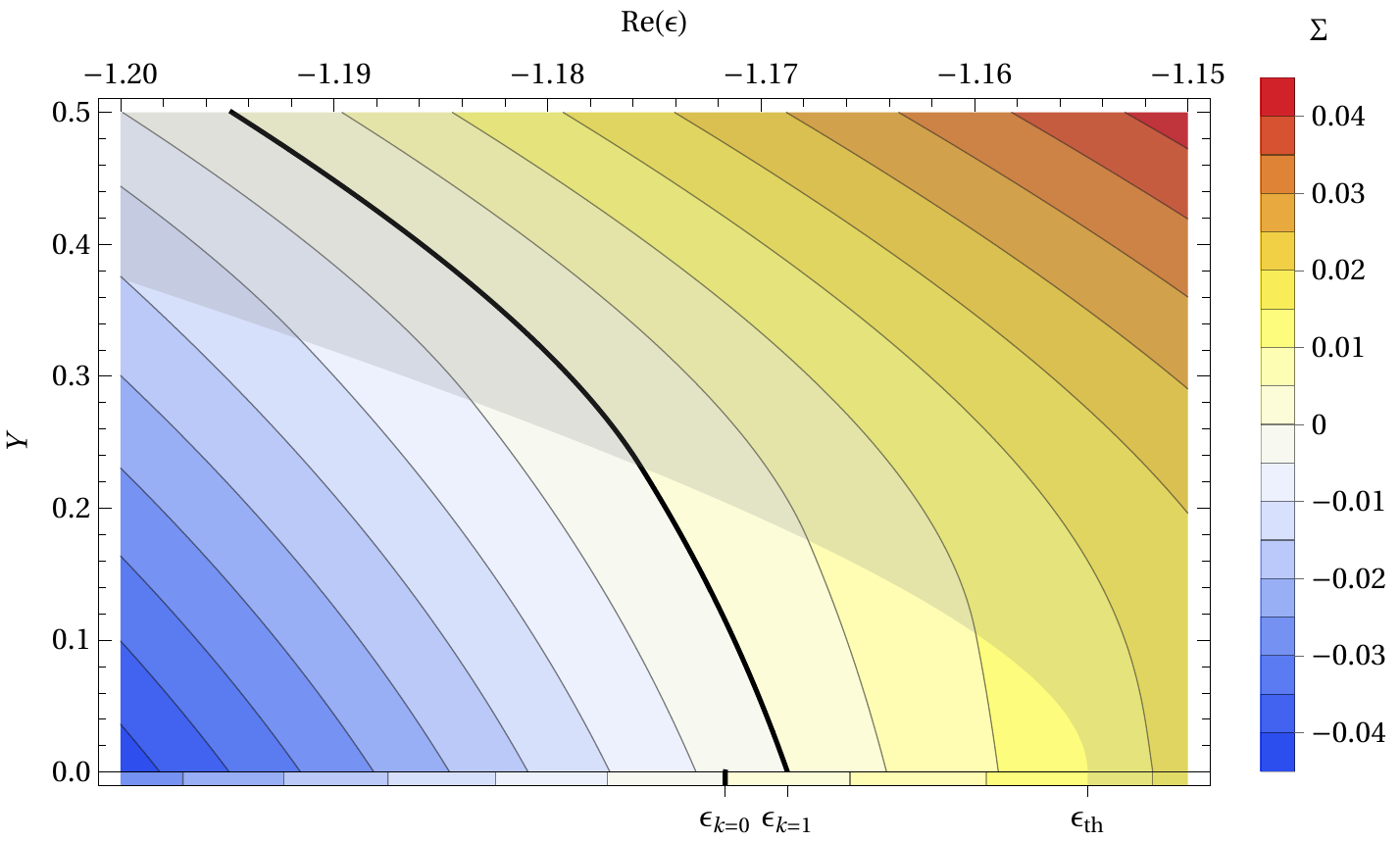}

  \caption{
    The complexity of the 3-spin spherical model in the complex plane, as a
    function of pure real energy and the magnitude $Y=\|\operatorname{Im}z\|^2/N$
    of the distance into the complex configuration space. The thick black
    contour shows the line of zero complexity, where stationary points become
    exponentially rare in $N$. The shaded region shows where stationary points
    have an ungapped spectrum. The complexity of the 3-spin model on the real
    sphere is shown below the horizontal axis; notice that it does not
    correspond with the limiting complexity in the complex configuration space
    below the threshold energy.
  } \label{fig:ground.complexity}
\end{figure}

Something more interesting is revealed if we zoom in on the complexity around
the ground state, shown in Fig.~\ref{fig:ground.complexity}. Here, the region
where most stationary points have a gapped hessian is shaded. The line
$\epsilon_\mathrm{gap}$ separating gapped from ungapped spectra corresponds
to the threshold energy $\epsilon_\mathrm{th}$ in the limit of $Y\to0$.  Above
the threshold, the limit of the complexity as $Y\to0$ (or
equivalently $r\to1$) also approaches the real complexity, plotted under the
horizontal axis. However, below the threshold this is no longer the case: here
the limit of $Y\to0$ of the complexity of complex
stationary points corresponds to the complexity $\Sigma_{k=1}$ of \emph{rank one saddles} in
the real problem, and their complexity becomes zero at $\epsilon_{k=1}$, where the
complexity of rank one saddles becomes zero \cite{Auffinger_2012_Random}.

There are several interesting features of the complexity. First is this
inequivalence between the real complexity and the limit of the complex
complexity to zero complex part. It implies, among other things, a desert of
stationary points in the complex plan surrounding the lowest minima, something
we shall see more explicitly in the next section. Second, there is only a small
collection of stationary points that appear with positive complexity and a
gapped spectrum: the small region in Fig.~\ref{fig:ground.complexity} that is
both to the right of the thick line and brightly shaded. We suspect that these
are the only stationary points that have any hope of avoiding participation
in Stokes points.

\subsection{Pure \textit{p}-spin: where are my neighbors?}
\label{subsec:p-spin.two.replica}

The problem of counting the density of Stokes points in an analytic
continuation of the spherical models is quite challenging, as the problem of
finding dynamic trajectories with endpoints at stationary points is already
difficult, and once made complex the problem has twice the number of fields
squared.

In this section, we begin to address the problem heuristically by instead
asking: if you are at a stationary point, where are your neighbors? The
stationary points geometrically nearest to a given stationary point should make
up the bulk of its adjacent points in the sense of being susceptible to Stokes
points. The distribution of these near neighbors in the complex configuration space therefore
gives a sense of whether many Stokes lines should be expected, and when.

To determine this, we perform the same Kac--Rice procedure as in the previous
section, but now with two probe points, or replicas, of the system. The simplify
things somewhat, we will examine the case where the only second probe is
complex; the first probe will be on the real sphere. The number of stationary
points with given energies $\epsilon_1\in\mathbb R$ and $\epsilon_2\in\mathbb C$ are, in the superfield formulation,
\begin{equation}
  \mathcal N^{(2)}
  =\int d\phi_1\,d\phi_2^*\,d\phi_2\,\exp\left\{
    \int d1 \left[
      \tilde\mathcal S_p(\phi_1(1))+\operatorname{Re}\tilde\mathcal S_p(\phi_2(1))
    \right]
  \right\}
\end{equation}
and we expect to find a two-spin complexity counting pairs of the form
\begin{equation}
  \Sigma^{(2)}=\lim_{N\to\infty}\frac1N\log\mathcal N^{(2)}
\end{equation}
which depends on the two energies and on mutual geometric invariants of the two probe points. The calculation follows exactly as before, but with an additional field. The average over $J$ is taken, and the supermatrix
\begin{equation}
  Q(1,2)
  =\left[
    \matrix{
      \phi_1(1)^T\phi_1(2) & \phi_1(1)^T\phi_2(2) & \phi_1(1)^T\phi_2(2)^* \cr
      \phi_2(1)^T\phi_1(2) & \phi_2(1)^T\phi_2(2) & \phi_2(1)^T\phi_2(2)^* \cr
      \phi_2(1)^\dagger\phi_1(2) & \phi_2(1)^\dagger\phi_2(2) & \phi_2(1)^\dagger\phi_2(2)^*
    }
  \right]
\end{equation}
is inserted with a Hubbard--Stratonovich transformation. The average number of pairs can then be written in the form
\begin{equation}
  \overline{\mathcal N^{(2)}}\propto\int dQ\,e^{NS_\mathrm{eff}[Q]}
\end{equation}
for the effective action
\begin{equation}
  \eqalign{
    S_\mathrm{eff}
    &=\int d1\,d2\,\operatorname{Tr}\left\{
      \frac14\left[\matrix{1&\frac12&\frac12\cr\frac12&\frac14&\frac14\cr\frac12&\frac14&\frac14}\right]Q^{(p)}(1,2)
      \right. \\
    &\qquad\qquad\left.-\frac p2\left[
      \matrix{\epsilon_1&0&0\cr0&\frac12\epsilon_2&0\cr0&0&\frac12\epsilon_2^*}
        \right]\left(
        Q(1,1)-I
      \right)\delta(1,2)
    \right\}+\frac12\det Q
  }
\end{equation}
Differentiating this with respect to $Q$, one finds the saddle point equations
\begin{equation}
  \eqalign{
    0=\frac{\partial S_\mathrm{eff}}{\partial Q(1,2)}
     &=
      \frac p4\left[\matrix{1&\frac12&\frac12\cr\frac12&\frac14&\frac14\cr\frac12&\frac14&\frac14}\right]\odot Q^{(p-1)}(1,2) \\
     &\qquad\qquad-\frac p2\left[
    \matrix{\epsilon_1&0&0\cr0&\frac12\epsilon_2&0\cr0&0&\frac12\epsilon_2^*}\right]\delta(1,2)
    +\frac12Q^{-1}(1,2)
  }
\end{equation}
where $\odot$ denotes element-wise multiplication. These are simplified by convolution to remove the superinverse, finally giving
\begin{equation}
  \eqalign{
    0
    &=\int d3\,\frac{\partial S_\mathrm{eff}}{\partial Q(1,3)}Q(3,2) \\
    &=\frac p4\int d3\,
      \left\{\left[\matrix{1&\frac12&\frac12\cr\frac12&\frac14&\frac14\cr\frac12&\frac14&\frac14}\right]\odot Q^{(p-1)}(1,3)\right\}Q(3,2) \\
                            &\qquad\qquad\qquad\qquad  -\frac p2 \left[
    \matrix{\epsilon_1&0&0\cr0&\frac12\epsilon_2&0\cr0&0&\frac12\epsilon_2^*}\right]Q(1,2)
    +\frac12I\delta(1,2)
  }
\end{equation}
Despite being able to pose the saddle point problem in a compact way, a great
deal of complexity lies within. The supermatrix $Q$ depends on 35 independent
bilinear products, and when the superfields are expanded produces 48 (not entirely independent) equations.
These equations can be split into 30 involving bilinear products of the
fermionic fields and 18 without them. The 18 equations without fermionic
bilinear products can be solved with a computer algebra package to eliminate 17
of the 20 non-fermionic bilinear products. The fermionic equations are
unfortunately more complicated.

They can be simplified somewhat by examination of the real two-replica problem.
There, all bilinear products involving fermionic fields from different
replicas, like $\eta_1^T\eta_2$, vanish. This is related to the influence of
the relative position of the two replicas to their spectra, with the vanishing
being equivalent to having no influence, i.e., the value of the determinant at
each stationary point is exactly what it would be in the one-replica problem
with the same invariants, e.g., energy and radius. Making this ansatz, the
equations can be solved for the remaining 5 bilinear products, eliminating all
the fermionic fields.

This leaves two bilinear products: $z_2^\dagger z_2$ and $z_2^\dagger z_1$, or one real and one complex number. The first is the radius
of the complex saddle, while the other is a complex generalization of the overlap. For us, it will be more convenient to work in terms of the
difference $\Delta z=z_2-z_1$ and the constants which characterize it, which
are $\Delta=\Delta z^\dagger\Delta z/N=\|\Delta z\|^2/N$ and
$\gamma=\frac{\Delta z^T\Delta z}{\|\Delta z\|}$. Once again we have one real
(and strictly positive) variable $\Delta$ and one complex variable $\gamma$.

\begin{figure}
  \hspace{5pc}
  \includegraphics{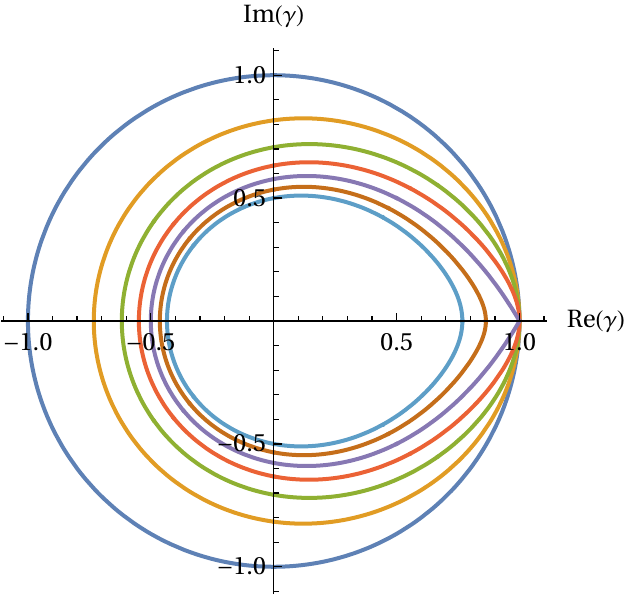}
  \hfill
  \includegraphics{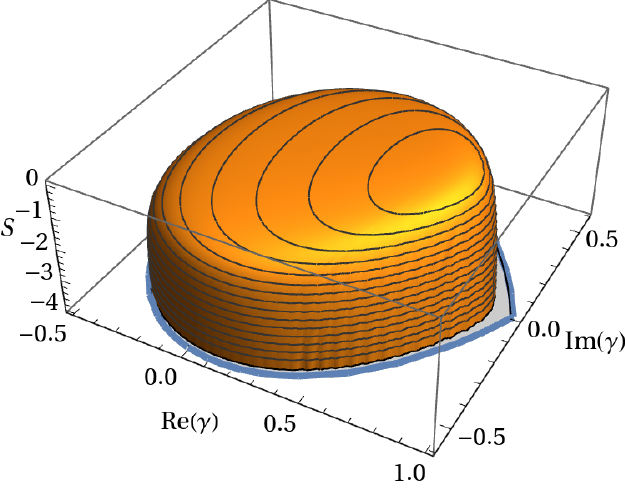}

  \caption{
    \textbf{Left:} The line bounding $\gamma$ in the complex plane as a function of
    $\Delta=0,1,2,\ldots,6$ (outer to inner). Notice that for $\Delta\leq4$,
    $|\gamma|=1$ is saturated for positive real $\gamma$, but is not for
    $\Delta>4$, and $\Delta=4$ has a cusp in the boundary. This is due to
    $\Delta=4$ corresponding to the maximum distance between any two points on
    the real sphere. \textbf{Right:} The two-spin complexity for $\Delta=4$ and
    some energy $\epsilon_1=\epsilon_2$. It approaches $-\infty$ at the boundary.
  } \label{fig:bound}
\end{figure}

Though the value of $\gamma$ is bounded  by $|\gamma|\leq1$ as a result of the
inequality $|\Delta z^T\Delta z|\leq\|\Delta z\|^2$, in reality this bound is not
the relevant one, because we are confined on the manifold $N=z^Tz$. The relevant
bound is most easily established by returning to a $2N$-dimensional real
problem, with $x=x_1$ and $z=x_2+iy_2$. The constraint gives $x_2^Ty_2=0$,
$x_1^Tx_1=1$, and $x_2^Tx_2=1+y_2^Ty_2$. Then, by their definitions,
\begin{equation}
  \Delta=1+x_2^Tx_2+y_2^Ty_2-2x_1^Tx_2=2(1+y_2^Ty_2-x_1^Tx_2)
\end{equation}
Define $\theta_{xx}$ as the angle between $x_1$ and $x_2$. Then $x_1^Tx_2=\|x_1\|\|x_2\|\cos\theta_{xx}=\sqrt{1-\|y_2\|}\cos\theta_{xx}$, and
\begin{equation}
  \Delta=2(1+\|y_2\|^2-\sqrt{1-\|y_2\|^2}\cos\theta_{xx})
\end{equation}
The definition of $\gamma$ likewise gives
\begin{equation}
  \eqalign{
    \gamma\Delta&=2-2x_1^Tx_2-2ix_1^Ty_2=2(1-\|x_2\|\cos\theta_{xx}-i\|y_2\|\cos\theta_{xy}) \\
          &=2(1-\sqrt{1-\|y_2\|^2}\cos\theta_{xx}-i\|y_2\|\cos\theta_{xy})
  }
\end{equation}
where $\theta_{xy}$ is the angle between $x_1$ and $y_2$.
There is also an inequality between the angles $\theta_{xx}$ and $\theta_{xy}$
between $x_1$ and $x_2$ and $y_2$, respectively, which takes that form
$\cos^2\theta_{xy}+\cos^2\theta_{xx}\leq1$. This results from the fact that
$x_2$ and $y_2$ are orthogonal, a result of the constraint.  These equations
along with the inequality produce the required bound on $|\gamma|$ as a
function of $\Delta$ and $\arg\gamma$, which is plotted in Fig.~\ref{fig:bound}.

A lot of information is contained in the full two-replica complexity, but we
will focus on the following question: what does the population of stationary
points nearby a given real stationary point look like? We think this is a
relevant question for the tendency for Stokes lines, for the following reason.
To determine whether two given stationary points, when tuned to have the same
imaginary energy, will share a Stokes line, one needs to solve what is known as
the global connection problem. As we have seen in \S\ref{subsec:stokes.conditions}, this as a question of a kind of
topological adjacency: two points will \emph{not} share a Stokes line if a third intervenes
with its thimble between them. We reason that the number of adjacent
stationary points of a given stationary point for a generic function in $D$
complex dimensions scales algebraically with $D$. Therefore, if the collection of
nearest neighbors has a nonzero complexity, i.e., scales \emph{exponentially}
with $D$, crowding around the stationary point in question, then these might be
expected to overwhelm the possible adjacencies, and so doing simplify the
problem of determining the properties of the true adjacencies. Until the
nonlinear flow equations are solved with dynamical mean field theory as has
been done for instantons \cite{Ros_2021_Dynamical}, this is the best heuristic.

For all displacements $\Delta$ and real energies
$\epsilon_1$, the maximum complexity is found for some real values of
$\epsilon_2$ and $\gamma$. Therefore we can restrict our study of the most
common neighbors to this. Note that the real part of $\gamma$ has a geometric
interpretation in terms of the properties of the neighbors: if a stationary
point sits in the complex configuration space near another,
$\operatorname{Re}\gamma$ can be related to the angle $\varphi$ made between
the vector separating these two points and the real configuration space as
\begin{equation}
  \varphi=\arctan\sqrt{\frac{1-\operatorname{Re}\gamma}{1+\operatorname{Re}\gamma}}
\end{equation}
Having concluded that the most populous neighbors are confined to real $\gamma$, we will make use of this angle instead of $\gamma$, which has a more direct geometric interpretation.

\begin{figure}
  \hspace{5pc}
  \includegraphics{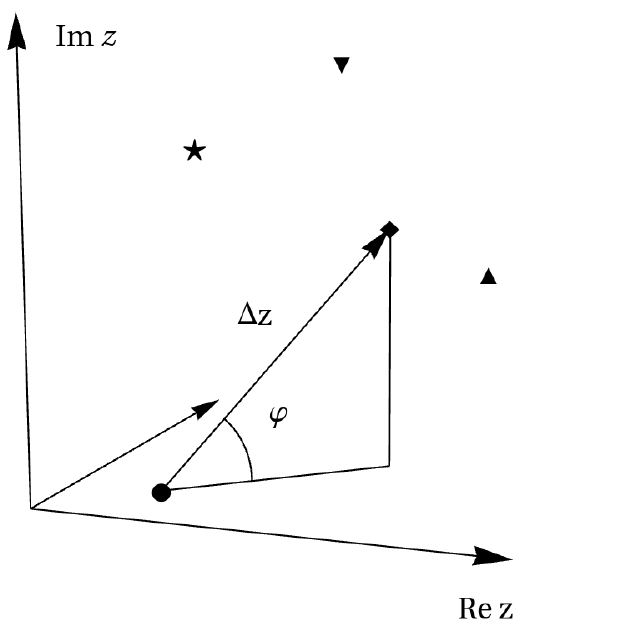}

  \caption{
    The geometric definition of the angle $\varphi$, between the displacement
    between two stationary points and the real configuration space.
  }
\end{figure}

\begin{figure}
  \includegraphics{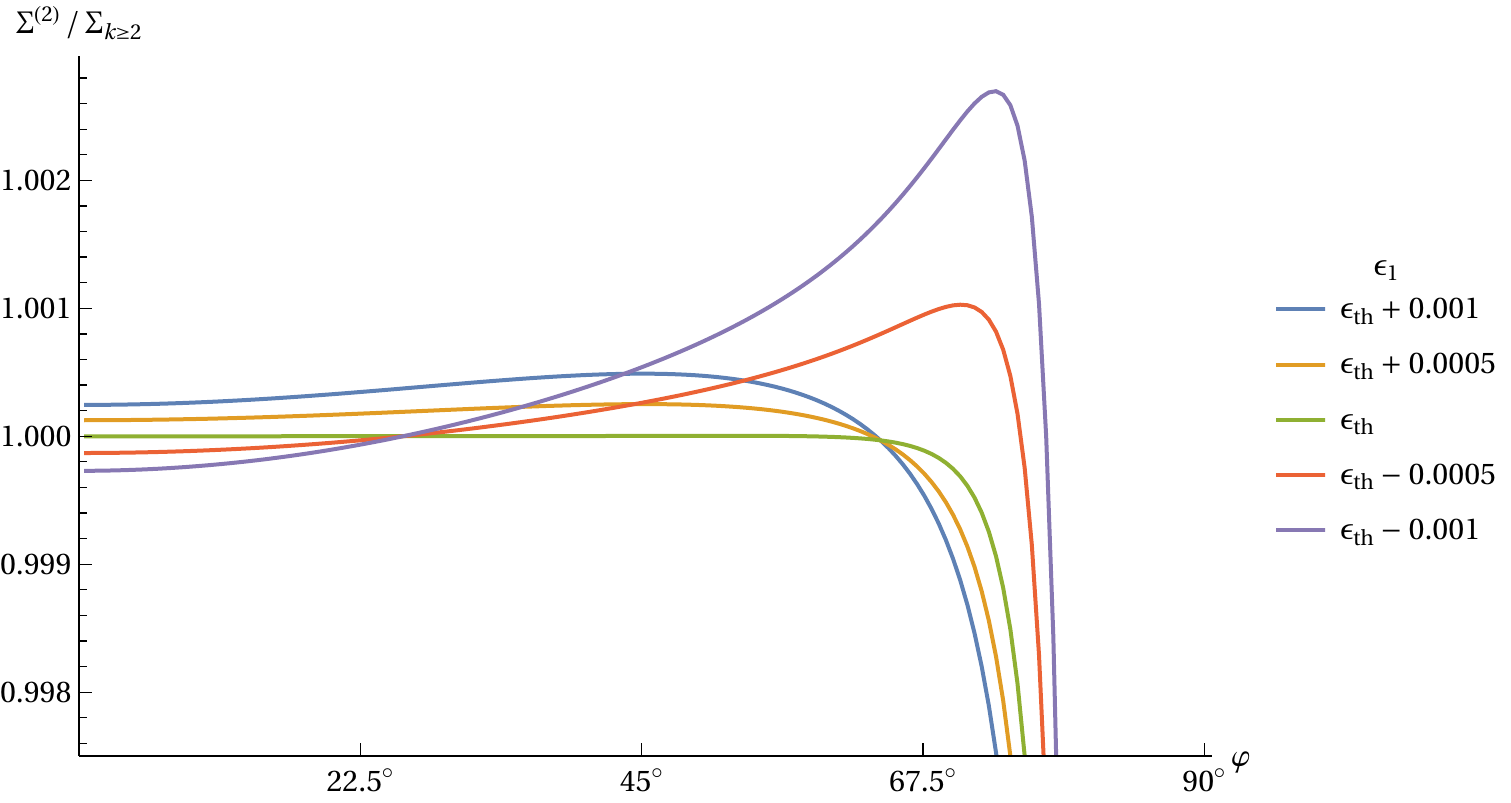}
  \caption{
    The scaled two-replica complexity $\Sigma^{(2)}$ as a function of angle
    $\varphi$ with $\epsilon_2=\epsilon_1$, $\Delta=2^{-7}$, and various
    $\epsilon_1$. At the threshold, the function undergoes a geometric
    transition and becomes sharper with decreasing $\Delta$.
  } \label{fig:neighbor.complexity.passing.threshold}
\end{figure}

First, we examine the importance of the threshold.
Fig.~\ref{fig:neighbor.complexity.passing.threshold} shows the two-replica
complexity evaluated at $\Delta=2^{-4}$ and equal energy
$\epsilon_2=\epsilon_1$ as a function of $\varphi$ for several $\epsilon_1$ as
the threshold is passed. The curves are rescaled by the complexity
$\Sigma_{k\geq2}(\epsilon_1)$ of index 2 and greater saddles in the real problem, which is what is
approached in the limit as $\Delta$ to zero. Below the threshold, the
distribution of nearby saddles with the same energy by angle is broad and
peaked around $\varphi=45^\circ$, while above the threshold it is peaked
strongly near the maximum $\varphi$ allowed by the bound. At the threshold, the function
becomes extremely flat.

One can examine the scaling of these curves as $\Delta$ goes to zero. Both
above and below the threshold, one finds a quickly-converging limit of
$(\Sigma^{(2)}/\Sigma_{k\geq2}-1)/\Delta$.
Above the threshold, these curves converge to a function whose peak is always
precisely at $45^\circ$, while below they converge to a function with a peak
that grows linearly with $\Delta^{-1}$ at $90^\circ$. At the threshold, the scaling is
different, and the function approaches a flat function extremely rapidly, as
$\Delta^3$.

\begin{figure}
  \includegraphics{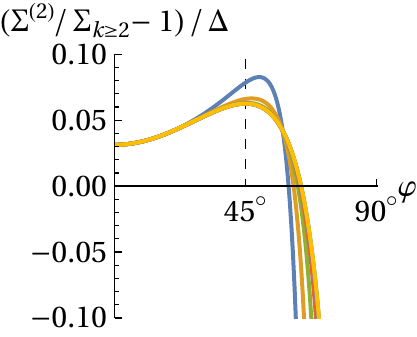}
  \hspace{-1em}
  \includegraphics{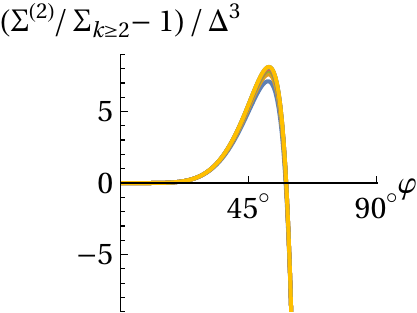}
  \hspace{-1em}
  \includegraphics{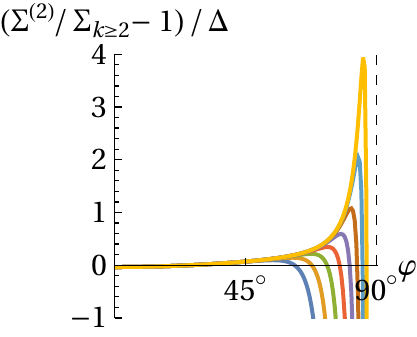}
  \hfill
  \includegraphics{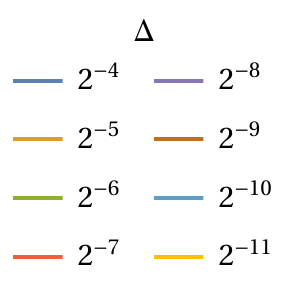}
  \caption{
    The scaled two-replica complexity $\Sigma^{(2)}$ as a function of angle
    $\varphi$ for various $\Delta$, $\epsilon_2=\epsilon_1$, and \textbf{Left:}
    $\epsilon_1=\epsilon_\mathrm{th}+0.001$ \textbf{Center:}
    $\epsilon_1=\epsilon_\mathrm{th}$ \textbf{Right:}
    $\epsilon=\epsilon_\mathrm{th}-0.001$. All lines have been normalized by
    the complexity $\Sigma_{k\geq2}$ of index 2 and greater saddles of the real 3-spin model.
  }
\end{figure}

Thus, there is an abrupt geometric transition in the population of nearest
neighbors as the threshold is crossed: above they are broadly distributed at
all angles, while below they are highly concentrated around $90^\circ$. From
this analysis it appears that the complexity of the nearest neighbors, at zero
distance, behaves as that of the index-2 saddles at all angles, which would
imply that the nearest neighbors vanish at the same point as the index-2
saddles. However, this is not the case: we have only shown that this is how the
neighbors at \emph{identical energy} scale, which is correct above the
threshold, but no longer underneath.

If an energy is taken under the threshold and the two-replica complexity
maximized with respect to both $\epsilon_2$ and $\varphi$, one finds that as
$\Delta\to0$, $\epsilon_2\to\epsilon_1$, as must be the case the find a
positive complexity at zero distance, but the maximum is never at
$\epsilon_2=\epsilon_1$, but rather at a small distance $\Delta\epsilon$ that
decreases with decreasing $\Delta$ like $\Delta^2$. When the complexity is
maximized in both parameters, one finds that, in the limit as $\Delta\to0$, the
peak is at $90^\circ$ but has a height equal to $\Sigma_{k=1}$, the complexity of
rank-1 saddles.

\begin{figure}
  \hfill\includegraphics{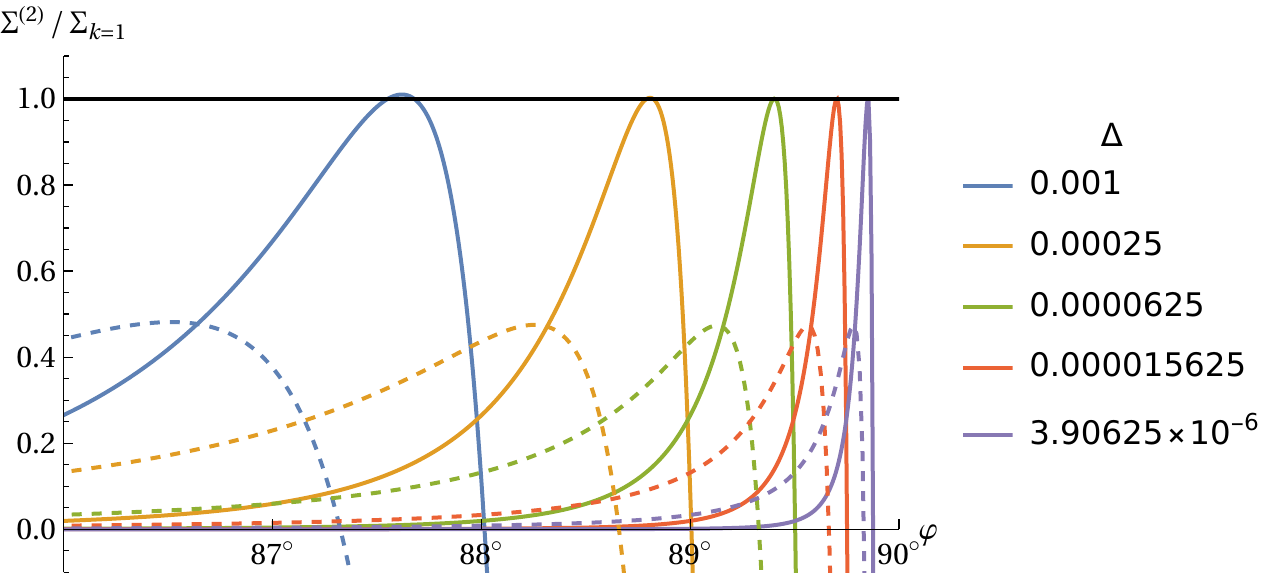}
  \caption{
    The two-replica complexity $\Sigma^{(2)}$ scaled by $\Sigma_{k=1}$ as a function of
    angle $\varphi$ for various $\Delta$ at $\epsilon_1=\epsilon_{k=2}$, the
    point of zero complexity for rank-two saddles in the real problem.
    \textbf{Solid lines:} The complexity evaluated at the value of $\epsilon_2$
    which leads to the largest maximum value. As $\Delta$ varies this varies
    like $\epsilon_2-\epsilon_1\propto\Delta^2$. \textbf{Dashed lines:} The
    complexity evaluated at $\epsilon_2=\epsilon_1$.
  }
\end{figure}

\begin{figure}
  \includegraphics{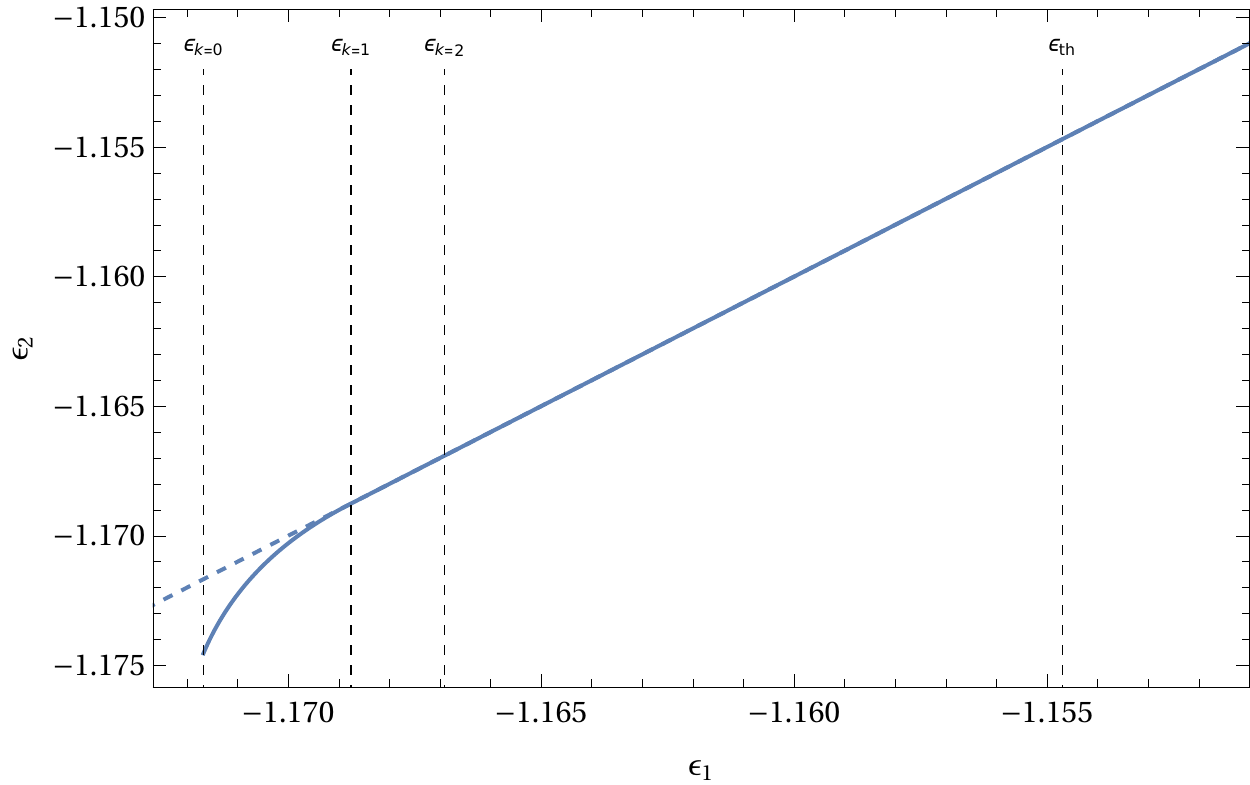}

  \caption{
    The energy $\epsilon_2$ of the nearest neighbor stationary points in the
    complex plane to a given real stationary point of energy $\epsilon_1$. The
    dashed line shows $\epsilon_2=\epsilon_1$. The nearest neighbor energy
    coincides with the dashed line until $\epsilon_{k=1}$, the energy where
    rank-one saddles vanish, where it peels off.
  } \label{fig:neighbor.energy}
\end{figure}

\begin{figure}
  \hfill\includegraphics{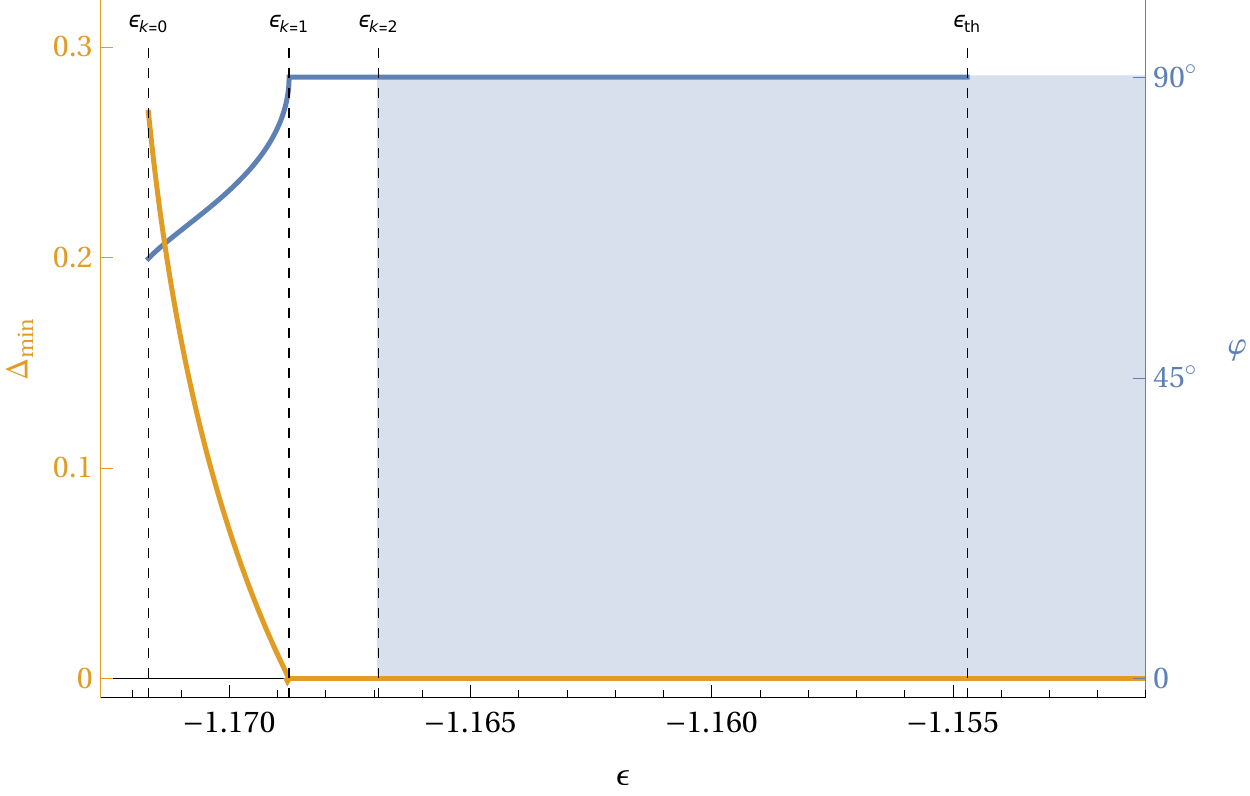}

  \caption{
    The properties of the nearest neighbor saddles in the 3-spin model as a function of energy
    $\epsilon$. Above the threshold energy $\epsilon_\mathrm{th}$, stationary
    points are found at arbitrarily close distance and at all angles $\varphi$
    in the complex plane. Below $\epsilon_\mathrm{th}$ but above $\epsilon_{k=2}$, stationary points are still found at arbitrarily close distance and
    all angles, but there are exponentially more found at $90^\circ$ than at
    any other angle. Below $\epsilon_{k=2}$ but above $\epsilon_{k=1}$, stationary
    points are found at arbitrarily close distance but only at $90^\circ$.
    Below $\epsilon_{k=1}$, neighboring stationary points are separated by a
    minimum squared distance $\Delta_\textrm{min}$, and the angle they are
    found at drifts. The complexity of nearest neighbors in the shaded region is $\Sigma_{k\geq2}$, while along the solid line for $\epsilon>\epsilon_{k=1}$ it is $\Sigma_{k=1}$. Below $\epsilon_{k=1}$ the complexity of nearest neighbors is zero.
  } \label{fig:nearest.properties}
\end{figure}

Below $\epsilon_{k=1}$, where the rank-1 saddle complexity vanishes, the complexity of stationary points of any type at zero distance is negative. To find what the nearest population looks like, one must find the minimum $\Delta$ at which the complexity is nonnegative, or
\begin{equation}
  \Delta_\textrm{min}=\operatorname{argmin}_\Delta\left(0\leq\max_{\epsilon_2, \varphi}\Sigma^{(2)}(\epsilon_1,\epsilon_2,\Delta,\varphi)\right)
\end{equation}
The result in $\Delta_\textrm{min}$ and the corresponding $\varphi$ that
produces it is plotted in Fig.~\ref{fig:nearest.properties}. As the energy is
brought below $\epsilon_{k=1}$, $\epsilon_2-\epsilon_1\propto
-|\epsilon_1-\epsilon_{k=1}|^2$, $\varphi-90^\circ\propto-|\epsilon_1-\epsilon_{k=1}|^{1/2}$, and $\Delta_\textrm{min}\propto|\epsilon_1-\epsilon_{k=1}|$. The
fact that the population of nearest neighbors has a energy lower than the
stationary point gives some hope for the success of continuation involving
these points: since Stokes points only lead to a change in weight when they
involve upward flow from a point that already has weight, neighbors that have a
lower energy won't be eligible to be involved in a Stokes line that causes a
change of weight until the phase of $\beta$ has rotated almost $180^\circ$. The
energy of nearest neighbors is plotted in Fig.~\ref{fig:neighbor.energy}, while
their angular distribution and distance is plotted in
Fig.~\ref{fig:nearest.properties}.

\subsection{Pure {\it p}-spin: numerics}
\label{subsec:p-spin.numerics}

To study Stokes lines numerically, we approximated them by parametric curves.
If $z_0$ and $z_1$ are two stationary points of the action with
$\operatorname{Re}\mathcal S(z_0)>\operatorname{Re}\mathcal S(z_1)$, then we
take the curve
\begin{equation}
  z(t)
  =(1-t)z_0+tz_1+(1-t)t\sum_{i=0}^mg_iP_i^{(1,1)}(2t-1)
\end{equation}
where the $g$s are undetermined complex vectors and the $P_i^{(1,1)}(x)$ are the
Jacobi polynomials, orthogonal on the interval $[-1,1]$ under the weight
$(1-x)(1+x)$. The Jacobi polynomials are used because they are orthogonal with
respect to integration over precisely the term they appear inside above. These
are fixed by minimizing a cost function, which has a global minimum only for
Stokes lines. Defining
\begin{equation}
  \mathcal L(t)
  = 1-\frac{\operatorname{Re}[\dot z(z(t))^\dagger z'(t)]}{|\dot z(z(t))||z'(t)|}
\end{equation}
where $\dot z(z)$ is the flow at $z$ given by \eqref{eq:flow}, this cost is given by
\begin{equation} \label{eq:cost}
  \mathcal C=\int_0^1 dt\,\mathcal L(t)
\end{equation}
$\mathcal C$ has minimum of zero, which is reached only by functions $z(t)$
whose tangent is everywhere parallel to the direction $\dot z$ of the dynamics.
Therefore, functions that satisfy $\mathcal C=0$ are time-reparameterized Stokes
lines.

We explicitly computed the gradient and hessian of $\mathcal C$ with respect to
the parameter vectors $g$. Stokes lines are found or not between points by
using the Levenberg--Marquardt algorithm starting from $g_i=0$ for all $i$,
and approximating the cost integral by a finite sum. To sample nearby
stationary points and assess their propensity for Stokes points, we do the
following. First, a saddle-finding routine based on Newton's method is run on
the \emph{real} configuration space of the $p$-spin model. Then, a
saddle-finding routine is run on the complex configuration space in the close
vicinity of the real saddle, using random initial conditions in a slowly
increasing radius of the real stationary point. When this process finds a new
distinct stationary point, it is finished. This method of sampling pairs heavily
biases the statistics we report here in favor of seeing Stokes points.

Once a pair of nearby stationary points has been found, one real and one in the
complex plane, their energies are used to compute the phase $\theta$ necessary
to give $\beta$ in order to set their imaginary energies to the same value, a
necessary condition for a Stokes line. A straight line (ignoring even the
constraint) is thrown between them and then minimized using the cost function
\eqref{eq:cost} for some initial $m=5$. Once a minimum is found, $m$ is
iteratively increased several times, each time minimizing the cost in between,
until $m=20$. If at some point in this process the cost blows up, indicating
that the solution is running away, the pair is thrown out; this happens
infrequently. At the end, there are several ways to asses whether a given
minimized line is a Stokes line: the value of the cost, the integrated
deviation from the constraint, and the integrated deviation from constant phase. Among minimized lines these values fall into
doubly-peaked histograms that well-separate prospective Stokes lines into
`good' and `bad' values for the given level of approximation $m$.

One cannot explicitly study the effect of crossing various landmark energies on
the $p$-spin in the system sizes that were accessible to our study, up to
around $N=64$, as the presence of, e.g., the threshold energy, is not noticeable until much larger size
\cite{Folena_2020_Rethinking}. However, we are
able to examine the effect of its symptoms: namely, the influence of the
spectrum of the stationary point in question on the likelihood that a randomly
chosen neighbor will share a Stokes line.

\begin{figure}
  \hfill\includegraphics{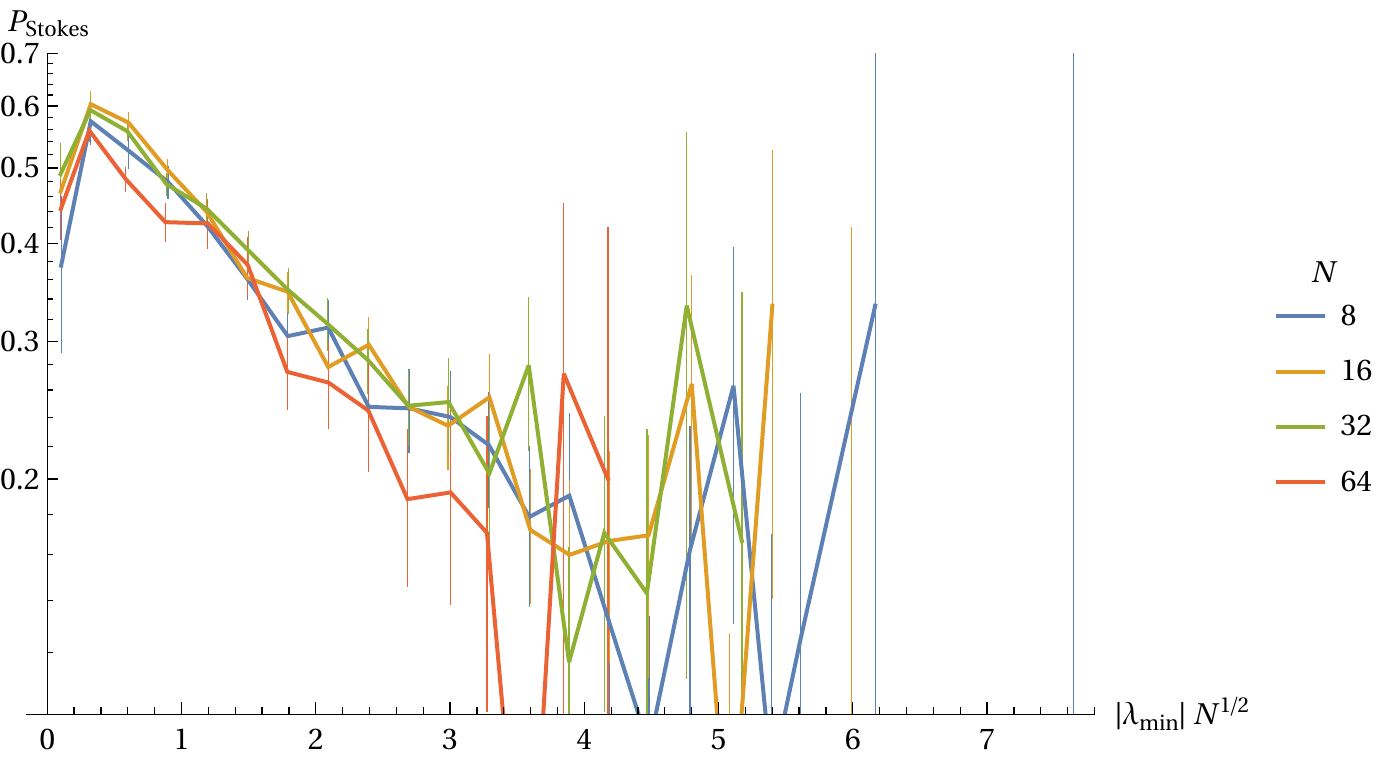}

  \caption{
    The probability $P_\mathrm{Stokes}$ that a real stationary point will share
    a Stokes line with its randomly chosen neighbor as a function of
    $|\lambda_\textrm{min}|$, the magnitude of the minimum eigenvalue of the
    hessian at the real stationary point. The horizontal axis has been rescaled
    to collapse the data at different system sizes $N$.
  } \label{fig:numeric.prob.eigenvalue}
\end{figure}

Data for the likelihood of a Stokes line as a function of the empirical gap
$|\lambda_\textrm{min}|$ of the real stationary point is shown in
Fig.~\ref{fig:numeric.prob.eigenvalue}. There, one sees that the probability of
finding a Stokes line with a near neighbor falls off as an exponential in the
magnitude of the smallest eigenvalue. As a function of system size, the tail
contracts like $N^{-1/2}$, which means that in the thermodynamic limit one
expects the probability of finding such a Stokes line will approach zero
everywhere expect where $\lambda_\textrm{min}\ll1$. This supports the idea that
gapped minima are unlikely to see Stokes lines.

\begin{figure}
  \hfill\includegraphics{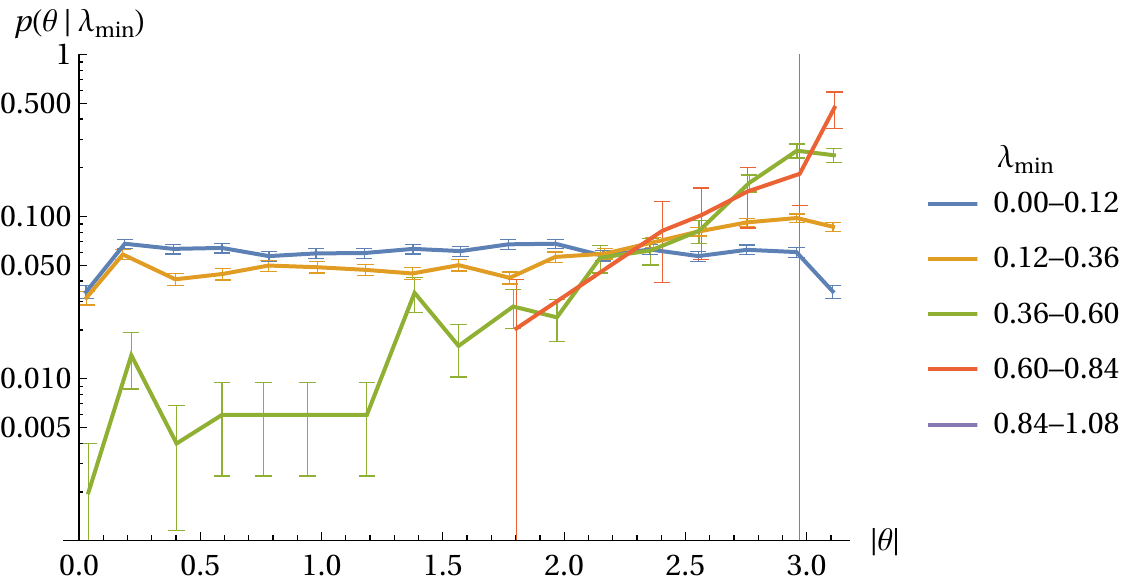}

  \caption{
    The probability density function for identified Stokes points as a function
    of $|\theta|$, the magnitude of the phase necessary to add to $\beta$ to
    reach the Stokes point, at $N=32$ and for several binned
    $|\lambda_\textrm{min}|$. As the empirical gap is increased, the population
    of discovered Stokes points becomes more concentrated around
    $|\theta|=\pi$.
  } \label{fig:numeric.angle.gap}
\end{figure}

We can also see that as the empirical gap is increased, Stokes points tend to
occur at very large phases. This can be seen for $N=32$ in
Fig.~\ref{fig:numeric.angle.gap}, which shows the probability distribution of
Stokes lines discovered as a function of phase $|\theta|$ necessary to reach
them. The curves are broken into sets representing different bins of the
empirical gap $|\lambda_\textrm{min}|$. As the empirical gap grows, Stokes
points become depleted around small phases and concentrate on very large ones.
This supports the idea that around the gapped minima, Stokes points will be
concentrated at phases that are nearly $180^\circ$, where the two-replica
calculation shows that almost all of their nearest neighbors will lie.

\subsection{Pure {\it p}-spin: is analytic continuation possible?}
\label{subsec:p-spin.continuation}

After all this work, one is motivated to ask: can analytic continuation be done in
even a simple complex model like the pure $p$-spin? Numeric and analytic
evidence indicates that the project is hopeless if ungapped stationary points
take a significant weight in the partition function, since for these Stokes
lines proliferate at even small continuation and there is no hope of tracking
them. However, for gapped stationary points we have seen compelling evidence
that suggests they will not participate in Stokes points, at least not until a
large phase rotation of the parameter being continued. This gives some hope for
continuation of the low-temperature thermodynamic phase of the $p$-spin, where
weight is concentrated in precisely gapped minima.

Recalling our expression \eqref{eq:real.thimble.partition.function} for the single-thimble contribution to the partition
function expanded to lowest order
in large $|\beta|$, we can write for
the $p$-spin after an infinitesimal rotation of $\beta$ into the complex plane
(before any Stokes points have been encountered)
\begin{equation}
  \eqalign{
    Z
    &=\sum_{\sigma\in\Sigma_0}n_\sigma Z_\sigma \\
    &\simeq\sum_{\sigma\in\Sigma_0}\left(\frac{2\pi}\beta\right)^{D/2}i^{k_\sigma}
      |\det\operatorname{Hess}\mathcal S(s_\sigma)|^{-\frac12}
      e^{-\beta\mathcal S(s_\sigma)} \\
    &\simeq\sum_{k=0}^D\int d\epsilon\,\mathcal N_\mathrm{typ}(\epsilon,k)
      \left(\frac{2\pi}\beta\right)^{D/2}i^k
      |\det\operatorname{Hess}\mathcal S(\epsilon,k)|^{-\frac12} e^{-\beta N\epsilon}
  }
\end{equation}
where $\mathcal N_\mathrm{typ}(\epsilon,k)$ is the typical number of stationary
points in a sample of the real $p$-spin model in the energy range $\epsilon$ to
$\epsilon+d\epsilon$ and with index $k$.
Following Derrida \cite{Derrida_1991_The}, this is related to the
\emph{average} number of stationary points in this range at large $N$ by
\begin{equation}
  \mathcal N_\mathrm{typ}(\epsilon,k)=\overline\mathcal N(\epsilon,k)+\eta(\epsilon,k)\overline\mathcal N(\epsilon,k)^{1/2}
\end{equation}
where $\eta$ is a random, sample-dependant number of order one. This gives two
terms to the typical partition function
\begin{equation}
  Z_\mathrm{typ}=Z_A+Z_B
\end{equation}
where
\begin{eqnarray} \fl
  Z_A
  \simeq\sum_{k=0}^D\int d\epsilon\,\overline\mathcal N(\epsilon,k)
    \left(\frac{2\pi}\beta\right)^{D/2}i^k
    |\det\operatorname{Hess}\mathcal S(\epsilon,k)|^{-\frac12} e^{-\beta N\epsilon}
    =\int d\epsilon\,e^{Nf_A(\epsilon)}
    \\ \fl
  Z_B
  \simeq\sum_{k=0}^D\int d\epsilon\,\eta(\epsilon,k)\overline\mathcal N(\epsilon,k)^{1/2}
    \left(\frac{2\pi}\beta\right)^{D/2}i^k
    |\det\operatorname{Hess}\mathcal S(\epsilon, k)|^{-\frac12}e^{-\beta N\epsilon} \\
    =\int d\epsilon\,\tilde\eta(\epsilon)e^{Nf_B(\epsilon)}
\end{eqnarray}
for functions $f_A$ and $f_B$ defined by
\begin{eqnarray}
  f_A
  &=-\beta\epsilon+\Sigma(\epsilon)-\frac12\int d\lambda\,\rho(\lambda\mid\epsilon)|\lambda|+\frac12\log\frac{2\pi}\beta
    +i\frac\pi2P(\lambda<0\mid\epsilon) \\
  f_B
  &=-\beta\epsilon+\frac12\Sigma(\epsilon)-\frac12\int d\lambda\,\rho(\lambda\mid\epsilon)|\lambda|+\frac12\log\frac{2\pi}\beta
    +i\frac\pi2P(\lambda<0\mid\epsilon)
\end{eqnarray}
and where $P(\lambda<0\mid\epsilon)$ is the cumulative probability distribution of the eigenvalues of the spectrum given $\epsilon$,
\begin{equation}
  P(\lambda<0\mid\epsilon)=\int_{-\infty}^0 d\lambda'\,\rho(\lambda'\mid\epsilon)
\end{equation}
and produces the macroscopic index $k/N$.
Each integral will be dominated by its value near the maximum of the real part of the exponential argument. Assuming that $\epsilon<\epsilon_\mathrm{th}$, this maximum occurs at
\begin{equation}
  0=\frac{\partial}{\partial\epsilon}\operatorname{Re}f_A\bigg|_{\epsilon=\epsilon_\mathrm{max}}=-\operatorname{Re}\beta-\frac12\frac{3p-4}{p-1}\epsilon_\mathrm{max}+\frac12\frac p{p-1}\sqrt{\epsilon_\mathrm{max}^2-\epsilon_\mathrm{th}^2}
\end{equation}
\begin{equation}
  0=\frac{\partial}{\partial\epsilon}\operatorname{Re}f_B\bigg|_{\epsilon=\epsilon_\mathrm{max}}=-\operatorname{Re}\beta-\epsilon_\mathrm{max}
\end{equation}
As with the 2-spin model, the integral over $\epsilon$ is oscillatory and can
only be reliably evaluated with a saddle point when either the period of
oscillation diverges \emph{or} when the maximum lies at a cusp. We therefore
expect changes in behavior when $\epsilon_\mathrm{max}=\epsilon_{k=0}$, the ground state energy.
The temperature at which this happens is
\begin{eqnarray}
  \operatorname{Re}\beta_A&=-\frac12\frac{3p-4}{p-1}\epsilon_{k=0}+\frac12\frac p{p-1}\sqrt{\epsilon_{k=0}^2-\epsilon_\mathrm{th}^2}\\
  \operatorname{Re}\beta_B&=-\epsilon_{k=0}
\end{eqnarray}
which for all $p\geq2$ has $\operatorname{Re}\beta_A\geq\operatorname{Re}\beta_B$.
Therefore, the emergence of zeros in $Z_A$ does not lead to the emergence of
zeros in the partition function as a whole, because $Z_B$ still produces a
coherent result (despite the unknown constant factor $\tilde\eta(\epsilon_{k=0})$). It is
only at $\operatorname{Re}\beta_B=-\epsilon_{k=0}$ where both terms contributing to
the partition function at large $N$ involve incoherent integrals near the
maximum, and only here where the density of zeros is expected to become
nonzero.

In fact, in the limit of $|\beta|\to\infty$, $\operatorname{Re}\beta_B$ is
precisely the transition found in \cite{Obuchi_2012_Partition-function} between
phases with and without a density of zeros. This value is an underestimate for
the transition for finite $|\beta|$, which likely results from the invalidity
of our large-$\beta$ approximation. More of the phase diagram might be
constructed by continuing the series for individual thimbles to higher powers
in $\beta$, which would be equivalent to allowing non-constant terms in the
Jacobian of the coordinate transformation over the thimble.

\begin{figure}
  \hspace{5pc}
  \hfill\includegraphics{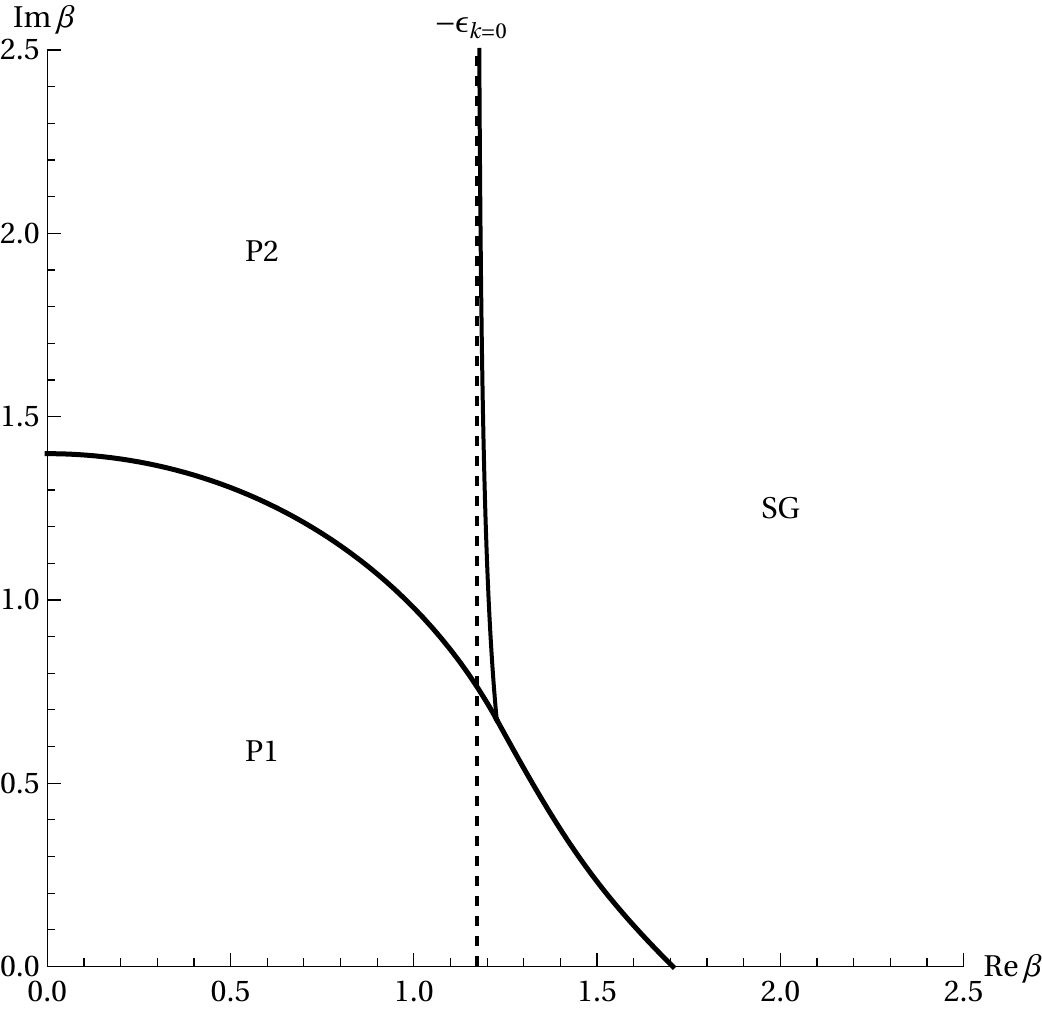}
  \caption{
    Phases of the 3-spin model in the complex-$\beta$, following Obuchi \&
    Takahashi \cite{Obuchi_2012_Partition-function}. The phase P2 contains a
    nonzero density of zeros of the partition function, while the `spin-glass'
    phase SG does not. Analytic continuation via thimbles correctly predicts
    the boundary between these two phases when $|\beta|\gg1$ to be
    $\operatorname{Re}\beta=-\epsilon_0$, shown with a dashed line.
  } \label{fig:obuchi_3-spin}
\end{figure}

This zeroth-order analysis for the $p$-spin suggests that analytic continuation
can be sometimes done despite the presence of a great many complex stationary
points. In particular, when weight is concentrated in certain minima Stokes
lines do not appear to interrupt the proceedings. How bad the situation is in
other regimes, like for smaller $|\beta|$, remains to be seen: our analysis
cannot tell between the effects of Stokes points changing the contour and the
large-$|\beta|$ saddle-point used to evaluate the thimble integrals. Taking the
thimbles to the next order in $\beta$ may reveal more explicitly where Stokes
points become important.

\section{Conclusion}
\label{sec:conclusion}

We have reviewed the Picard--Lefschetz technique for analytically continuing
integrals and examined its applicability to the analytic continuation of configuration
space integrals over the pure $p$-spin models. The evidence suggests that
analytic continuation is possible when weight is concentrated in gapped minima,
who seem to avoid Stokes points, and is likely intractable otherwise.

This has implications for the ability to analytically continue other types of
theories. For instance, \emph{marginal} phases of glasses, spin glasses, and
other problems are characterized by concentration in pseudogapped minima. Based
on the considerations of this paper, we suspect that analytic continuation is
never tractable in such a phase, as Stokes points will always proliferate among
even the lowest minima.

It is possible that a statistical theory of analytic continuation could be
developed in order to treat these cases, whereby one computes the average or
typical rate of Stokes points as a function of stationary point properties, and
treats their proliferation to complex saddles as a structured diffusion
problem. This would be a very involved calculation, involving counting
classical trajectories with certain boundary conditions, but in principle it
could be done as in \cite{Ros_2021_Dynamical}. Here the scale of the
proliferation may rescue things, allowing accurate statements to be
made about its average effect.

\section*{References}
\bibliographystyle{unsrt}
\bibliography{stokes}

\end{document}